\documentclass[journal]{IEEEtran}
\usepackage{cite}
\usepackage{url}
\usepackage{graphicx}
\usepackage{color}
\usepackage{placeins}
\usepackage{float}
\usepackage{tabularx,colortbl}
\usepackage{ifthen}
\usepackage{caption}
\usepackage{amsmath}
\usepackage{bm}
\usepackage{amssymb}
\usepackage{amsfonts}
\usepackage{xcolor}
\usepackage{titlesec}
\usepackage{tocloft}
\usepackage{tikz}
\usetikzlibrary{shapes.geometric, arrows, positioning}
\usepackage{afterpage}  
\usepackage{xcolor}
\usepackage{stfloats} 
\usepackage{comment}
\usetikzlibrary{decorations.pathreplacing}
\usepackage{tikz}
\usetikzlibrary{calc, decorations.pathreplacing}
\usepackage{graphicx}
\usepackage{subcaption}
\usepackage{caption}
\usepackage{amsmath, amssymb}
\usepackage{graphicx}
\usepackage{bm}
\usepackage{enumitem}
\usepackage{algorithm}
\usepackage{algpseudocode}
\usepackage{booktabs}
\usepackage{multirow}
\usepackage{booktabs}

\title{Multi-User MIMO Enhancement using Metasurface Wavefront Bending (MWB)}

\author{Xiaolu Yang, Oscar C\'{e}spedes Vicente, and Christophe Caloz,~\IEEEmembership{Fellow,~IEEE}%
\thanks{X. Yang, O. C\'{e}spedes Vicente, and C. Caloz are with the
Department of Electrical Engineering (ESAT), KU Leuven,
Kasteelpark Arenberg 10, 3001 Leuven, Belgium
(e-mail: xiaolu.yang@kuleuven.be).}%
}

\newcommand{\pati}[1]{}

\begin{document}
\maketitle
\begin{abstract}
This paper introduces metasurface wavefront bending (MWB) to enhance spatial multiplexing in radiative near-field multi-user multiple-input multiple-output (MU-MIMO) systems. By increasing spherical-wave curvature, MWB strengthens range-dependent phase variations across the receiver array, reduces inter-user channel correlation and improves user separability. The paper develops a curvature-dependent channel analysis, a generalized-sheet-transition-condition (GSTC) synthesis procedure for MWB and a complete metasurface-assisted MU-MIMO channel model. Both a practical common-profile scheme and an ideal user-specific benchmark are evaluated. The results demonstrate that MWB provides substantial improvements in spectral efficiency and effective channel rank. Finally, a three-layer transmissive Huygens metasurface architecture is proposed for physical implementation.
\end{abstract}

\vskip0.5\baselineskip


\section{Introduction}
\label{sec:Introduction}
\pati{Background}
Future wireless communication systems are expected to operate at increasingly high carrier frequencies and to employ extremely large antenna apertures~\cite{marzetta2010noncooperative,bjornson2024towards,lu2024tutorial,cui2022near,koutsos2022analysis}. Since the far-field limit distance scales as $d_{\mathrm{F}}\approx 2D^{2}/\lambda$~\cite{balanis2016antenna}, where $D$ denotes the largest antenna array dimension and $\lambda$ the wavelength, the combined increase in aperture size ($D$)\footnote{The physical size of a simple antenna (e.g., dipole antenna) scales down with frequency (e.g., $\ell = \lambda/2 = c/(2f)$). However, modern wireless communication systems use antenna \emph{arrays}, where only the size of the radiative elements (typically patches) decreases with frequency, whereas the overall array size ($D$) may remain identical for larger directive gain---from larger electrical size---or even increase in absolute terms, for super-high directive gain.} and decrease in wavelength ($\lambda$) will expand the radiative near-field zone. Regions that were in the far field in earlier systems may fall within the near field in future deployments. Within these regions,  the impinging electromagnetic fields are no longer plane waves over the entire receiver aperture. Instead, they exhibit spherical wavefronts, whose spatial phase variation across the receiver apertures depends not only on the angle of arrival but also on the transmitter range~\cite{lu2022communicating, ramezani2023exploiting}. This range-dependent spatial structure provides an additional degree of freedom for distinguishing users located in similar angular directions but at different distances, thereby creating an opportunity for enhancing spatial multiplexing in near-field multi-user multiple-input multiple-output (MU-MIMO) systems.

\pati{Research Gap}
Most near-field MIMO studies to date have focused on characterizing, estimating or processing channels naturally determined by the propagation geometry of the system. Representative research directions include extremely large array channel modeling~\cite{lu2022communicating}, near-field beam focusing~\cite{zhang2022beam}, near-field multiple access~\cite{wu2023multiple}, channel estimation~\cite{cui2022channel}, and near-field RIS beamforming~\cite{bjornson2021primer,mei2022study,papazafeiropoulos2024near,ataloglou2025reconfigurable}. These studies have established that curved, typically spherical, phase fronts can enhance channel capacity. However, they generally treat the curved electromagnetic wavefronts incident on the receiver array as a consequence of the underlying propagation conditions. They neither elucidate the fundamental physics underlying this effect nor explore the corresponding opportunities for electromagnetic engineering.

\pati{Contributions}
In this work, we fill up this gap, by providing the electromagnetic explanation of the capacity enhancement in terms of physical reduced channel correlation, and using properly engineered metasurfaces~\cite{holloway2012overview, minatti2016synthesis,di2020smart,achouri2021electromagnetic,wani2021thin} to reshape the wireless environment for higher multiplexing gain. In the latter, source-shifting transmissive metasurfaces increases the effective curvature of an incident spherical wavefront, which we shall refer to as \emph{metasurface wavefront bending (MWB)}, so as to enhance near-field MU-MIMO multiplexing. The corresponding transformation is used as a design prescription and the metasurfaces are synthesized using the generalized sheet transition conditions (GSTCs)~\cite{achouri2015general}, which relate the desired field discontinuities to the required electric and magnetic surface susceptibilities. The resulting transmitted wavefronts exhibit increased effective curvature on the receiver side, thereby strengthening the range-dependent spatial phase variation across the base-station array, reducing inter-user channel correlation, and enhancing spatial multiplexing capability. The paper further develops a complete metasurface-assisted electromagnetic channel model, evaluates both a practical common-profile curvature-bending scheme and an ideal user-specific benchmark, and proposes a three-layer transmissive Huygens (electric and magnetic) metasurface architecture for physical implementation~\cite{pfeiffer2013metamaterial,epstein2016huygens,achouri2021electromagnetic}.

\pati{Organization}
The remainder of the paper is organized as follows. Section~\ref{sec:principle} explains the fundamental principle of MWB through an intuitive phase-difference and phasor-addition interpretation. Section~\ref{sec:model_withoutMS} formulates the free-space near-field uplink MU-MIMO model and shows how spherical-wave curvature affects normalized channel correlation. Section~\ref{sec:MS_GSTC} presents the GSTC-based metasurface synthesis used to realize the curvature-bending transformation. Section~\ref{sec:model_withMS} integrates the synthesized transformation into a complete user--metasurface--base-station channel model. Section~\ref{sec:performance_withMS} evaluates common-profile MWB and the user-specific benchmark. Section~\ref{sec:implementation} presents a physical implementation based on a three-layer transmissive Huygens metasurface. Finally, Sec.~\ref{sec:conclusion} concludes the paper.

\section{Principle}
\label{sec:principle}
\begin{figure}[h!]
\centering
\includegraphics[width=9cm]{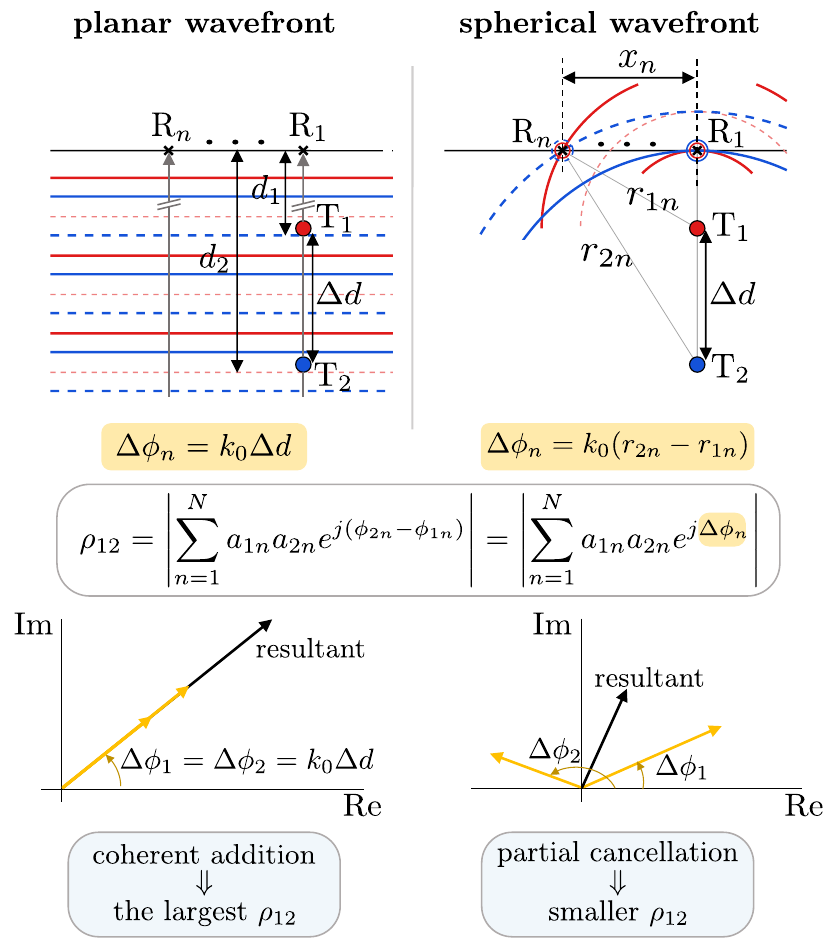}
\caption{Channel correlation reduction from planar to spherical wavefronts.}
\label{fig:visual_correlation}
\end{figure}

\pati{Setup and Correlation Definition}
Figure~\ref{fig:visual_correlation} highlights a fundamental difference between planar and spherical wavefronts in terms of channel correlation. Consider two transmitters, $T_1$ and $T_2$, located at different ranges from a receiver array, with axial distances $d_1$ and $d_2$, respectively, and range separation $\Delta d=d_2-d_1$.
The receiver array consists of $N$ elements, denoted by $R_1,\ldots,R_n,\ldots,R_N$. The corresponding channel vectors are written as $\mathbf h_1=[h_{11},\ldots,h_{1N}]^{\mathrm T}$ and $\mathbf h_2=[h_{21},\ldots,h_{2N}]^{\mathrm T}$, with elements $h_{1n}=a_{1n}e^{j\phi_{1n}}$ and $h_{2n}=a_{2n}e^{j\phi_{2n}}$, where $\phi_{1n}$ and $\phi_{2n}$ denote the phases produced by $T_1$ and $T_2$, respectively, at the $n$th element of the receiver, leading to the relative phase difference $\Delta\phi_n = \phi_{2n}-\phi_{1n}$. The correlation between the two channel vectors is then given by~\cite{bjornson2017massive}
\begin{equation}
\rho_{12}
=|\mathbf h_1^{\mathrm H}\mathbf h_2|
=\left|\sum_{n=1}^{N}h_{1n}^{*}h_{2n}\right|
=\left|\sum_{n=1}^{N}a_{1n}a_{2n}e^{j\Delta\phi_n}\right|,
\label{eq:principle_channel_correlation}
\end{equation}
which may be interpreted as a vector sum of phasors, where the $n$-th phasor has the magnitude $a_{1n}a_{2n}$ and angle $\Delta\phi_n$. 

\pati{Correlation Reduction from Planar to Spherical Fronts}
In the case \emph{planar} wavefronts, represented in the left of Fig.~\ref{fig:visual_correlation}, the path-length difference between the waves radiated by the two users is $\Delta d$ at all the receiver elements, i.e.,  $\Delta\phi_n=k_0\Delta d$ is constant. Therefore, all phasors in Eq.~\eqref{eq:principle_channel_correlation} point in the same direction and add coherently, yielding a maximum resultant magnitude for the given phasor amplitudes. By contrast, for \emph{spherical} wavefronts, shown in the right panel of Fig.~\ref{fig:visual_correlation}, the relative phase difference depends on the receiver-element position. Denoting the propagation distances from $T_1$ and $T_2$ to the $n$th receiver element as $r_{1n}$ and $r_{2n}$, respectively, the relative phase difference is $\Delta\phi_n = k_0\left(r_{2n}-r_{1n}\right)$, and denoting the transverse coordinate of the $n$th receiver element as $x_n$, the propagation distances can be approximated as
\begin{subequations}
\begin{align}
r_{1n}&=\sqrt{d_1^2+x_n^2}\approx d_1+\frac{x_n^2}{2d_1},
\label{eq:spheri_r1n}\\
r_{2n}&=\sqrt{d_2^2+x_n^2}\approx d_2+\frac{x_n^2}{2d_2}.
\label{eq:eq:spheri_r2n}
\end{align}
\end{subequations}
The resulting phase difference is then
\begin{equation}
\Delta \phi_n = k_0 \left(r_{2n}-r_{1n}\right)\approx k_0\left[\Delta d+\left(\frac{1}{2d_2}-\frac{1}{2d_1}\right)x_n^2\right],
\end{equation}
which is a function of $n$, causing the phasors in Eq.~\eqref{eq:principle_channel_correlation} to point in different directions and partially cancel. This partial cancellation reduces the magnitude of their resultant and, hence, the channel correlation.

\begin{figure}[h!]
\centering
\includegraphics[width=8cm]{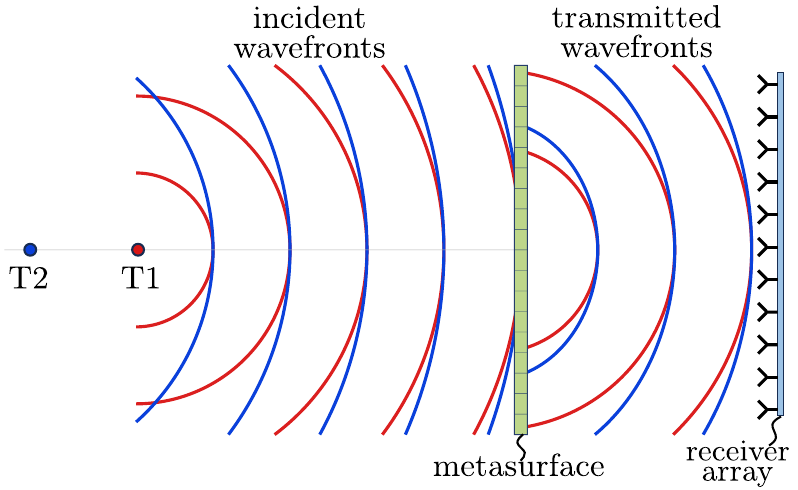}
\caption{Illustration of wavefront bending (WFB) by a properly designed metasurface placed between the transmitters and the receiver array to reduce correlation between transmitters as shown in Fig.~\ref{fig:visual_correlation}.}
\label{fig:visual_MWB}
\end{figure}

\begin{figure}[h!]
\centering
\includegraphics[width=8cm]{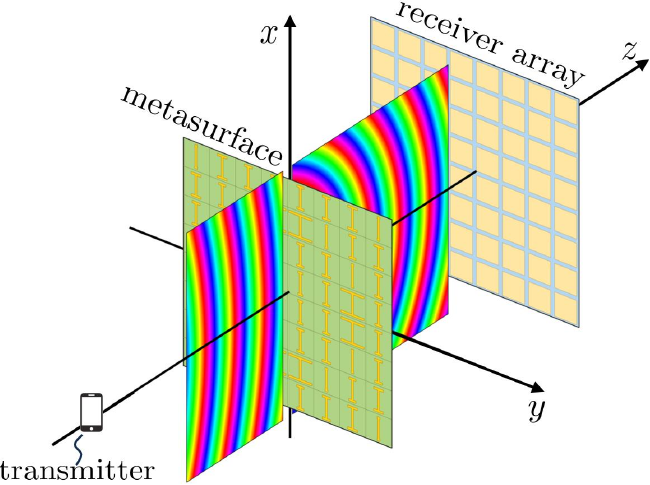}
\caption{Perspective representation of the WFB metasurface in Fig.~\ref{fig:visual_MWB} for a single transmitter on.}
\label{fig:principle_system}
\end{figure}

\pati{Metasurface wavefront Bending}
The preceding discussion demonstrates that spherical-wave curvature can reduce channel correlation by producing an inter-user phase difference that varies across the receiver aperture. This observation motivates the deliberate engineering of wavefront curvature illustrated in Fig.~\ref{fig:visual_MWB}. A transmissive metasurface may be designed to impose a curvature-bending transformation on the incident spherical wavefronts, thereby increasing the effective curvature of the received fields. The resulting enhancement in spatial phase variation across the receiver array is expected to improve the distinguishability of the user channels, which will be demonstrated later. Figure~\ref{fig:principle_system} shows, for later use, a perspective view of Fig.~\ref{fig:visual_MWB} with wavefront bending (WFB) for a single user on.

%

\section{Curvature-Dependent MU-MIMO}
\label{sec:model_withoutMS}

\subsection{Channel Model}
\begin{figure}[h!]
\centering
\includegraphics[width=8cm]{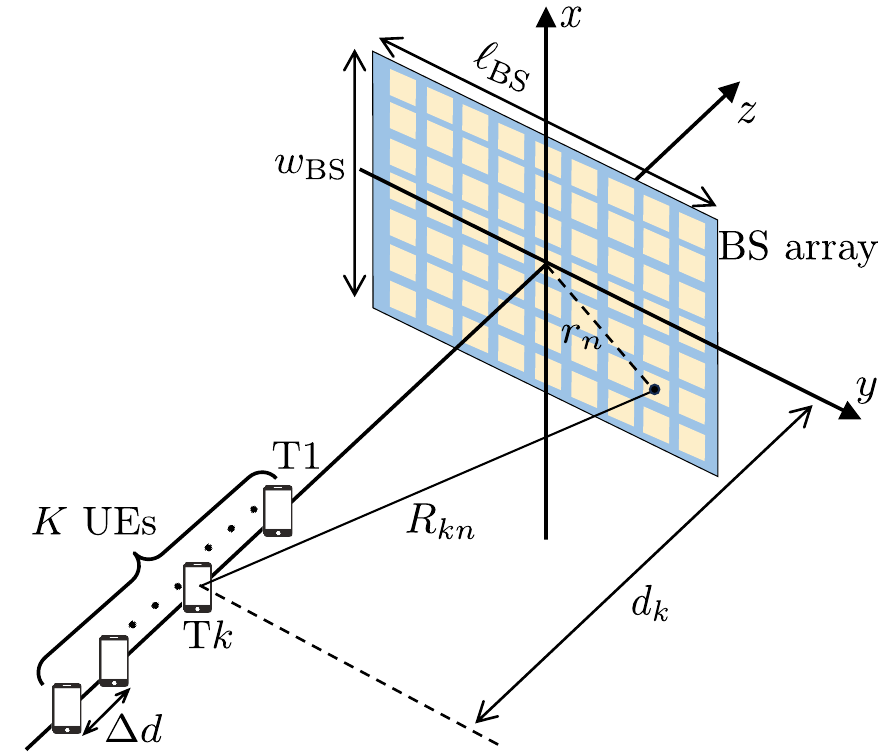}
\caption{Uplink MU-MIMO system without metasurface.}
\label{fig:config_withoutMS}
\end{figure}

\pati{System Configuration}
We consider an uplink communication system comprising $K$ single-antenna user equipments (UEs) and a base-station (BS) receiver array with $N$ antenna elements, as illustrated in Fig.~\ref{fig:config_withoutMS}. The BS array is positioned in the $x$--$y$ plane, with its center selected as the origin of a Cartesian coordinate system. The position of the $n$th BS element is denoted by $\mathbf r_n=(x_n,y_n,0)$, with $n=1,\ldots,N$. To isolate the effect of wavefront curvature, we consider the most-challenging configuration, that shown in the figure, where the UEs are aligned in a row perpendicular to the BS and differ only in range. UE $k$ is located at $\mathbf s_k=(0,0,-d_k)$, with $d_k=d_1+(k-1)\Delta d$, where $d_k$ denotes the axial distance from UE $k$ to the center of the BS array and $\Delta d$ denotes the range spacing between adjacent UEs. This configuration provides no angular separability, so user discrimination relies solely on the spherical-wave curvature observed across the aperture, as illustrated in Fig.~\ref{fig:visual_correlation}.

%

\pati{Channel Model}
The symbol transmitted by UE $k$ is modeled as $s_k\sim\mathcal{CN}(0,p_k)$, where $\mathcal{CN}$ denotes the circularly symmetric complex Gaussian distribution\footnote{A circularly symmetric complex Gaussian random variable has a probability distribution that remains unchanged under any rotation in the complex plane. Specifically, if $s_k\sim\mathcal{CN}(0,p_k)$, then $e^{j\theta}s_k=s_k$ for any $\theta\in[0,2\pi)$. Equivalently, writing $s_k=s_{k,\mathrm R}+j s_{k,\mathrm I}$, its real and imaginary parts are independent and satisfy $s_{k,\mathrm R},s_{k,\mathrm I}\sim \mathcal{N}(0,p_k/2)$. Consequently, $\mathbb{E}\{|s_k|^2\}=p_k$.}~\cite{tse2005fundamentals}, with $0$ being the mean of and $p_k$ the variance, which corresponds to the transmit power of UE $k$. The received signal vector at the BS may then be written as
\begin{equation}
\mathbf y=\sum_{k=1}^{K}\mathbf h_k s_k+\mathbf n,
\label{eq:rx_signal_vector}
\end{equation}
where $\mathbf h_k=[h_{k1},h_{k2},\ldots,h_{kN}]^{\mathrm T}\in\mathbb C^N$, with the superscript $T$ denoting the transpose operation, represents the channel vector from UE $k$ to the BS array. The vector $\mathbf n\sim\mathcal{CN}(\mathbf 0,\sigma^2\mathbf I_N)$ represents circularly symmetric complex Gaussian noise\footnote{The distribution $\mathbf n\sim\mathcal{CN}(\mathbf 0,\sigma^2\mathbf I_N)$ indicates that $\mathbb{E}\{\mathbf n\}=\mathbf 0$ and $\mathbb{E}\{\mathbf n\mathbf n^{\mathrm H}\} =\sigma^2\mathbf I_N$. Therefore, each noise component satisfies $n_n\sim\mathcal{CN}(0,\sigma^2)$ and has average power $\mathbb{E}\{|n_n|^2\}=\sigma^2$.}. The received signal at the $n$th BS element is therefore
\begin{equation}
y_n=\sum_{k=1}^{K}h_{kn}s_k+n_n .
\label{eq:rx_signal_element}
\end{equation}
For line-of-sight free-space propagation, the channel coefficient between UE $k$ and BS element $n$ corresponds to the scalar spherical Green's function~\cite{balanis2016antenna}
\begin{equation}
h_{kn}
=C_0\frac{e^{-jk_0R_{kn}}}{R_{kn}},
\label{eq:sw_channel_withoutMS}
\end{equation}
where $C_0=jk_0^2/(4\pi\omega\epsilon_0)$ for a dipole source and
\begin{equation}
R_{kn}=\|\mathbf r_n-\mathbf s_k\|
=\sqrt{x_n^2+y_n^2+d_k^2}
=\sqrt{r_n^2+d_k^2},
\label{eq:Rkn_withoutMS}
\end{equation}
is the propagation distance from UE $k$ to BS element $n$.

\pati{Detection}
To detect UE $k$, the BS multiplies the received signal vector $\mathbf y$ [Eq.~\eqref{eq:rx_signal_vector}] by a combining vector $\mathbf v_k\in\mathbb C^N$, which yields
\begin{align}
\hat{s}_k
&=\mathbf v_k^{\mathrm H}\mathbf y
=\mathbf v_k^{\mathrm H}\mathbf h_k s_k
+\underbrace{\sum_{i\neq k}\mathbf v_k^{\mathrm H}\mathbf h_i s_i}_{\text{interference}}
+\underbrace{\mathbf v_k^{\mathrm H}\mathbf n}_{\text{noise}},
\label{eq:combined_signal}
\end{align}
with the superscript H denoting the the Hermitian conjugate operation and where the successive terms correspond to the desired signal, residual multi-user interference and receiver noise, respectively. In this operation, the BS uses the minimum mean-square-error (MMSE) combining vector
\begin{equation}
\mathbf v_k^{\mathrm{MMSE}}
=\left(\sum_{i=1}^{K}p_i\mathbf h_i\mathbf h_i^{\mathrm H}
+\sigma^2\mathbf I_N\right)^{\!\!\!-1}\mathbf h_k,
\label{eq:MMSE_combiner}
\end{equation}
that maximizes the desired-signal against interference and noise~\cite{madhow1994mmse}. 

\pati{General Performance Metrics}
For a general $K$-user system, one may quantify the interference of UE $k$ from the other $K-1$ users as noise and define the corresponding signal-to-interference-plus-noise ratio (SINR) as~\cite{bjornson2017massive}
\begin{equation}
\mathrm{SINR}_k
=\frac{p_k|\mathbf v_k^{\mathrm H}\mathbf h_k|^2}
{\sum_{i\neq k}p_i|\mathbf v_k^{\mathrm H}\mathbf h_i|^2
+\sigma^2\|\mathbf v_k\|^2},
\label{eq:SINR_general}
\end{equation}
which corresponds to the spectral efficiency
\begin{equation}
\eta_k=\log_2(1+\mathrm{SINR}_k).
\label{eq:SE}
\end{equation}
We may also define the normalized channel correlation between pair of UEs~$i$ and~$k$ among the $K$ users as
\begin{equation}
\bar{\rho}_{ik}
=\frac{|\mathbf h_i^{\mathrm H}\mathbf h_k|}
{\|\mathbf h_i\|\|\mathbf h_k\|}
=
\frac{\left|\sum_{n=1}^{N}h_{in}^{*}h_{kn}\right|}
{\sqrt{\sum_{n=1}^{N}|h_{in}|^2}
\sqrt{\sum_{n=1}^{N}|h_{kn}|^2}},
\label{eq:normalized_rho}
\end{equation}
which measures the spatial separability between these UEs independently of their path loss, with value close to one indicating very poor separability and value close to zero indicating nearly orthogonal channels.

To characterize the joint spatial separability of all the $K$ users, we consider the channel matrix ${\mathbf H}=\{\mathbf h_1,\mathbf h_2,\ldots,\mathbf h_K\}$. Let $\{\sigma_i\}_{i=1}^{r}$, where $r$ is the rank of ${\mathbf H}$, denote the list of nonzero singular values of the normalized channel matrix $\widetilde{\mathbf H}=\mathbf H/||\mathbf H||$ and define the corresponding normalized weights as $q_i=\sigma_i/\sum_{j=1}^{r}\sigma_j$. We can then define the effective rank~\cite{roy2007effective} the channel matrix as
\begin{equation}
r_{\mathrm{eff}}
=\exp\left(-\sum_{i=1}^{r}q_i\ln q_i\right),
\label{eq:effective_rank}
\end{equation}
which approaches one when the $K$ channels are nearly parallel and approaches $K$ when they are nearly orthogonal. 

\pati{Performance Metrix for $K=2$}
For the particular case $K=2$, the SINR of UE~$k$ in Eq.~\eqref{eq:SINR_general} can be expressed more intuitively in terms of its interference-free SNR,
\begin{equation}
\gamma_k
\triangleq
\frac{p_k}{\sigma^2}.
\label{eq:reference_snr}
\end{equation}
The corresponding SINR formula is obtained by substituting Eq.~\eqref{eq:MMSE_combiner} into Eq.~\eqref{eq:SINR_general}, replacing the channel vectors with their normalized counterparts, $\widetilde{\mathbf h}_k\triangleq\mathbf h_k/\|\mathbf h_k\|$, and identifying the quantities defined in Eqs.~\eqref{eq:normalized_rho} and~\eqref{eq:reference_snr} in the resulting expression, which yields
\begin{equation}
\overline{\mathrm{SINR}}_{k}
=\gamma_k\left(1-\frac{\gamma_k}{1+\gamma_k}\bar{\rho}_{ik}^{\,2}\right),
\label{eq:sinr_vs_correlation}
\end{equation}
whose complete derivation is provided in Appendix~\ref{app:sinr_correlation_derivation}. 

\subsection{Curvature-Dependent Performance for Two-User MIMO}
\label{sec:correlation_vs_curvature}

\pati{Motivation}
We shall consider here the simplest case, $K=2$, as in Fig.~\ref{fig:visual_correlation}, to show how spherical-wave curvature affects channel multiplexing performance. This case suffices to reveal the relationship between wavefront curvature and spatial separability in a direct and intuitive manner. The subsequent analysis of the metasurface-assisted system extends the performance evaluation to general $K$-user scenarios.

\pati{Curvature Definition}
For the spherical wavefront generated by UE $k$, the propagation distance $R_{kn}$ determines the local curvature~\cite{do2016differential} at the $n$th BS element as
\begin{equation}
\kappa_{kn}
=\frac{1}{R_{kn}}
=\frac{1}{\sqrt{d_k^2+r_n^2}},
\label{eq:local_curvature}
\end{equation}
where $r_n^2=x_n^2+y_n^2$ (Fig.~\ref{fig:config_withoutMS}). At the center of the BS array, where $r_n=0$, this curvature reduces to
\begin{equation}
\frac{1}{d_k}\triangleq \kappa_k.
\label{eq:center_curvature}
\end{equation}
Substituting Eq.~\eqref{eq:center_curvature} into Eq.~\eqref{eq:local_curvature} gives
\begin{equation}
\kappa_{kn}
=\frac{\kappa_k}{\sqrt{1+\kappa_k^2r_n^2}},
\qquad
\frac{\partial \kappa_{kn}}{\partial \kappa_k}
=\frac{1}{(1+\kappa_k^2r_n^2)^{3/2}}>0.
\label{eq:local_curvature_from_center}
\end{equation}
where the first equation relates the local curvature~$\kappa_{kn}$ to the central curvature~$\kappa_k$, while the second shows that~$\kappa_{kn}$ increases monotonically with~$\kappa_k$, so that~$\kappa_k$ may be used as a unique parameter to quantify the overall curvature.

\pati{Curvature-Dependent Performance}
Considering two UEs at distances $d_k$ and $d_i$ from the receiver array and with separation $\Delta d=d_i-d_k$. The curvature of UE $i$ is related to that of UE $k$ from Eq.~\eqref{eq:center_curvature}, as
\begin{equation}
\kappa_i=\frac{1}{d_i}=\frac{1}{d_k+\Delta d}
=\frac{\kappa_k}{1+\kappa_k\Delta d}.
\label{eq:kappa_relation}
\end{equation}
We can then evaluate the channel correlation between the two users in term of curvature. For this purpose, we first explicit Eq.~\eqref{eq:normalized_rho} with Eq.~\eqref{eq:sw_channel_withoutMS}, which  yields
\begin{equation}
\bar{\rho}_{ik}
=\frac{\left|\sum_{n=1}^{N}\dfrac{e^{-jk_0(R_{kn}-R_{in})}}{R_{in}R_{kn}}\right|}
{\sqrt{\sum_{n=1}^{N}\dfrac{1}{R_{in}^2}}\sqrt{\sum_{n=1}^{N}\dfrac{1}{R_{kn}^2}}},
\label{eq:exact_norm_rho}
\end{equation}
where we approximate the phase terms as
\begin{align}
k_0R_{kn}=k_0\left(\sqrt{d_k^2+r_n^2}\right)
&\approx k_0\left(d_k+\frac{r_n^2}{2d_k}\right),
\label{eq:fresnel_phase_expansion}
\end{align}
the amplitude terms as $1/R_{kn}\approx1/d_k$, and substitute Eq.~\eqref{eq:kappa_relation} in the resulting expression, which leads to the normalized channel correlation
\begin{equation}
\bar{\rho}_{ik}
\approx
\left|\frac{1}{N}\sum_{n=1}^{N}\exp\left[-j\frac{k_0}{2}\frac{\kappa_k^2\Delta d}{1+\kappa_k\Delta d}r_n^2\right]\right|,
\label{eq:rho_vs_curvature}
\end{equation}
as a function of the central curvature~$\kappa_k$. Assuming~$\Delta d>0$, the factor~$\kappa_k^2\Delta d/(1+\kappa_k\Delta d)$ increases monotonically with~$\kappa_k$. Therefore, a larger curvature produces greater phase variation across the BS elements through the factor~$r_n^2$, resulting in stronger phasor cancellation and, in general, lower channel correlation, as illustrated in Fig.~\ref{fig:visual_correlation} in the extreme comparison between planar wavefronts ($\kappa_k=0$) and spherical wavefronts ($\kappa_k>0$).

Figure~\ref{fig:performance_vs_curvature} presents the curvature-dependent performance for the two-user uplink MU-MIMO system. Figure~\ref{fig:performance_vs_curvature}(a) plots, using Eq.~\eqref{eq:exact_norm_rho}, $\bar{\rho}_{ik}$ as a function of $\kappa_k$ for several values of $\Delta d$. As $\kappa_k$ approaches zero, the correlation approaches unity because the wavefronts are locally planar over the BS aperture. As $\kappa_k$ increases, the greater phase variation across the aperture reduces the correlation. In addition, for a fixed $\kappa_k$, a larger range separation $\Delta d$ leads to a lower correlation.
Figure~\ref{fig:performance_vs_curvature}(b) shows the corresponding SINR of UE~$k$, evaluated using Eq.~\eqref{eq:sinr_vs_correlation} with an interference-free reference SNR of $\gamma_k=10~\mathrm{dB}$. In the plane-wave limit, $\kappa_k\to 0$ and $\bar{\rho}_{ik}\to 1$, Eq.~\eqref{eq:sinr_vs_correlation} gives $\overline{\mathrm{SINR}}_{k}=10\left(1-{10}/{11}\right)={10}/{11}$, which corresponds to $-0.41~\mathrm{dB}$. As the curvature increases, the reduction in $\bar{\rho}_{ik}$ enables more effective suppression of co-user interference, and the SINR gradually approaches the interference-free reference value of $10~\mathrm{dB}$.
\begin{figure}[h!]
    \centering
    \begin{subfigure}[t]{0.49\linewidth}
        \includegraphics[width=\linewidth]{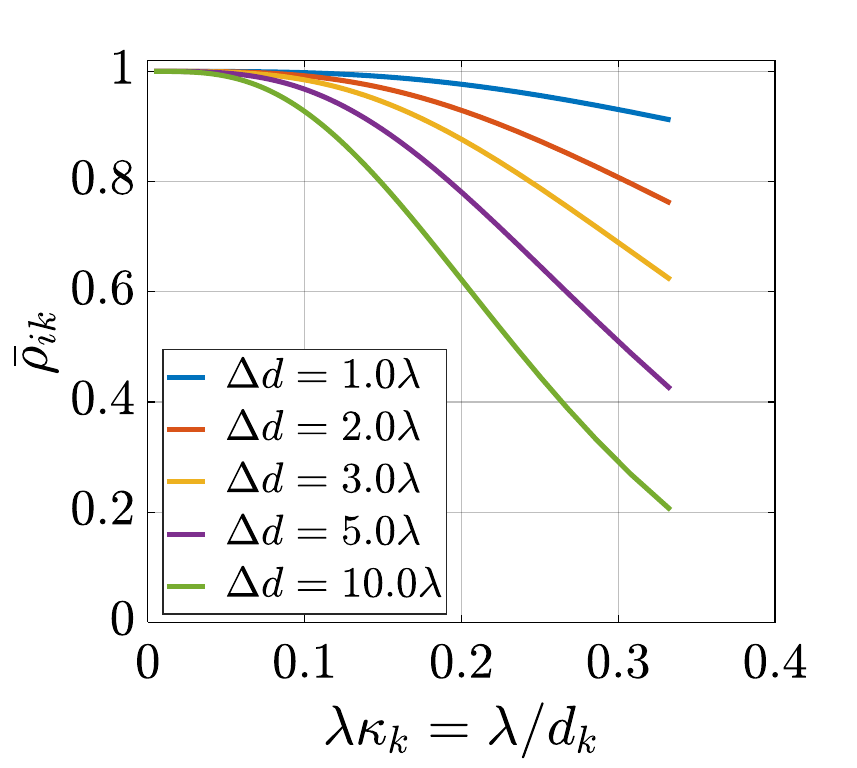}
        \captionsetup{skip=1pt}
        \caption{}
    \end{subfigure}
    \hfill
    \begin{subfigure}[t]{0.49\linewidth}
        \includegraphics[width=\linewidth]{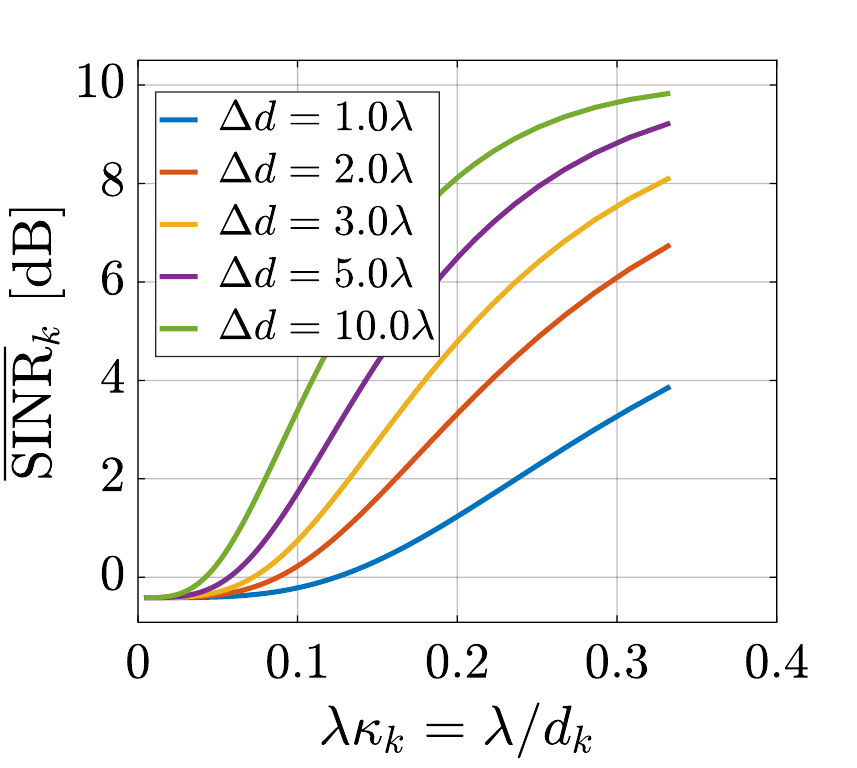}
        \captionsetup{skip=1pt}
        \caption{}
        \label{fig:performance_vs_curvature}
    \end{subfigure}
\caption{Curvature-dependent performance for two-user MIMO with different user spacings. (a)~Normalized channel correlation $\bar{\rho}_{ik}$ [Eq.~\eqref{eq:exact_norm_rho}] versus central curvature $\lambda\kappa_k$. (b)~SINR of UE $k$ [Eq.~\eqref{eq:sinr_vs_correlation}] versus $\lambda\kappa_k$, evaluated using an interference-free reference SNR of $\gamma_k=10~\mathrm{dB}$. The system configuration corresponds to Fig.~\ref{fig:config_withoutMS}, with a $10\times10$ BS array and element spacing of $\lambda/2$, corresponding to aperture dimensions of $w_{\mathrm{BS}}=\ell_{\mathrm{BS}}=4.5\lambda$. The operating frequency is $f=10~\mathrm{GHz}$.}
\label{fig:performance_vs_curvature}
\end{figure}
%
\section{Metasurface wavefront Bending}
\label{sec:MS_GSTC}

\begin{figure}[!h]
    \centering
    \begin{subfigure}[t]{0.49\linewidth}
        \includegraphics[width=\linewidth]{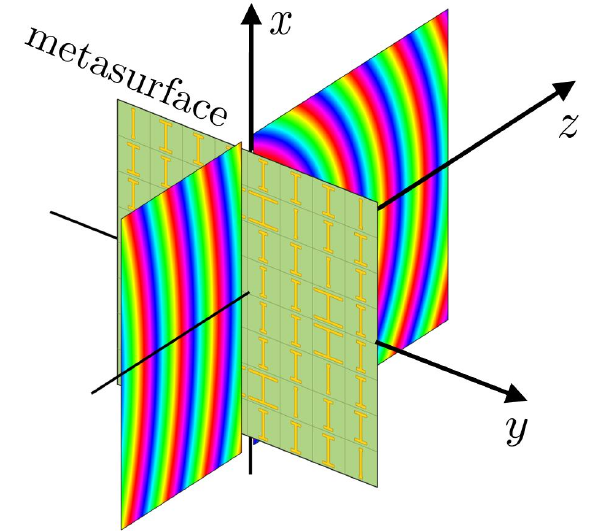}
        \captionsetup{skip=1pt}
        \caption{}
    \end{subfigure}
    \hfill
    \begin{subfigure}[t]{0.49\linewidth}
        \includegraphics[width=\linewidth]{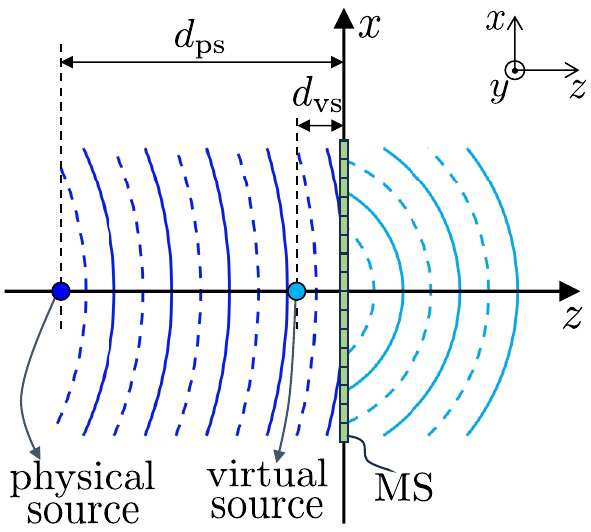}
        \captionsetup{skip=1pt}
        \caption{}
    \end{subfigure}
\caption{Metasurface wavefront bending (WFB) transformation for a single user. (a)~Perspective illustration of the transformation. The metasurface is represented schematically because its deeply subwavelength unit-cell details are not visible at this scale. (b)~Corresponding longitudinal-sectional view ($x$-$z$ plane), with physical source at distance $d_{\mathrm{ps}}$ from the metasurface and virtual source for the transmitted field located at the shorter distance~$d_{\mathrm{vs}}$.}
\label{fig:specif_transformation}
\end{figure}

\pati{Motivation}
The benefits of increased wavefront curvature demonstrated in Sec.~\ref{sec:correlation_vs_curvature} motivate the use of a WFB metasurface, conceptually illustrated with the receiver array in Figs.~\ref{fig:visual_MWB} and~\ref{fig:principle_system} and repeated with technical details in Fig.~\ref{fig:specif_transformation}.

\pati{Specified Incident Field}
A dipolar physical source generates an incident spherical wavefront, which the WFB metasurface transforms into a transmitted wavefront with greater curvature, as conceptually illustrated in Fig.~\ref{fig:specif_transformation}(a). As shown in Fig.~\ref{fig:specif_transformation}(b), this transformation is achieved by converting the incident field produced by the physical source in the plane of the metasurface, at the distance $d_{\mathrm{ps}}$,
\begin{subequations}
\begin{align}
\mathbf{E}_{\mathrm{tan}}^{-}(x,y)
&=\mathbf{E}_{\mathrm{tan}}^{\mathrm{dip}}\left(x,y;d_{\mathrm{ps}}\right),\\
\mathbf{H}_{\mathrm{tan}}^{-}(x,y)
&=\mathbf{H}_{\mathrm{tan}}^{\mathrm{dip}}\left(x,y;d_{\mathrm{ps}}\right),
\end{align}
\label{eq:incident_field_ms}
\end{subequations}
into a new field associated with a nearby virtual source, placed at the shorter distance $d_{\mathrm{vs}}$. Exact expressions of the dipole fields in Eqs.~\eqref{eq:incident_field_ms} are available in textbooks~\cite{balanis2016antenna}.

\pati{Specified Transmitted Field}
Prescribing the desired transmitted field to exhibit the wavefront associated with the virtual source and the same magnitude as the incident field, we have
\begin{subequations}
\begin{align}
\mathbf{E}_{\mathrm{tan}}^{+}(x,y)
&=\left|\mathbf{E}_{\mathrm{tan}}^{-}(x,y)\right|\exp\left[j\angle \mathbf{E}_{\mathrm{tan}}^{\mathrm{dip}}\left(x,y;d_{\mathrm{vs}}\right)\right],
\\
\mathbf{H}_{\mathrm{tan}}^{+}(x,y)
&=\left|\mathbf{H}_{\mathrm{tan}}^{-}(x,y)\right|\exp\left[j\angle \mathbf{H}_{\mathrm{tan}}^{\mathrm{dip}}\left(x,y;d_{\mathrm{vs}}\right)\right],
\end{align}\label{eq:target_field_ms}
\end{subequations}
where the superscripts $-$ and $+$ denote the fields immediately before and after the metasurface, i.e., at $z=0^-$ and $z=0^+$, respectively. According to Eq.~\eqref{eq:center_curvature}, the distances $d_{\mathrm{ps}}$ and $d_{\mathrm{vs}}$ produce the wavefront curvatures of $1/d_{\mathrm{ps}}$ and $1/d_{\mathrm{vs}}$, respectively, at the metasurface center. The condition $d_{\mathrm{vs}}<d_{\mathrm{ps}}$ therefore produces curvature amplification, which we quantify through the bending ratio
\begin{equation}
\beta=\frac{d_\text{ps}}{d_\text{vs}}>1 .
\label{eq:bending_ratio_def}
\end{equation}
%

\pati{Corresponding GSTC Equations}
To determine the metasurface required to realize the transformation prescribed in Eq.~\eqref{eq:target_field_ms}, we relate the electromagnetic fields immediately before and after the metasurface through GSTCs~\cite{achouri2015general}. Assuming a tangential monoanisotropic metasurface, the GSTCs read
\begin{subequations}
\begin{align}
\hat{\mathbf z}\times\Delta\mathbf H
&=
j\omega\epsilon_0\overline{\overline{{\chi}}}_{\mathrm{ee}}\mathbf E_{\mathrm{av}},\\
\Delta\mathbf E\times\hat{\mathbf z}
&=
j\omega\mu_0\overline{\overline{{\chi}}}_{\mathrm{mm}}\mathbf H_{\mathrm{av}},
\end{align}\label{eq:GSTC_tensor}
\end{subequations}
where $\overline{\overline{\chi}}_{\mathrm{ee}}$ and $\overline{\overline{\chi}}_{\mathrm{mm}}$ denote the electric and magnetic surface susceptibility tensors, respectively, and where we define the field jumps and averages as
\begin{subequations}\label{eq:field_avg_jump}
\begin{align}
\Delta\mathbf E
&=\mathbf E^{+}-\mathbf E^{-},&\qquad\Delta\mathbf H
&=\mathbf H^{+}-\mathbf H^{-},\\
\mathbf E_{\mathrm{av}}
&=\frac{\mathbf E^{+}+\mathbf E^{-}}{2},&\qquad\mathbf H_{\mathrm{av}}
&=\frac{\mathbf H^{+}+\mathbf H^{-}}{2}.
\end{align}
\end{subequations}
Inserting Eqs.~\eqref{eq:field_avg_jump} into Eqs.~\eqref{eq:GSTC_tensor} yields the scalar equations
\begin{subequations}
\begin{align}
-\Delta H_y
&=j\omega\epsilon_0
\left(\chi_\text{ee}^{xx}E_{x,\mathrm{av}}+
\chi_\text{ee}^{xy}E_{y,\mathrm{av}}\right),
\label{eq:gstc_scalar_1}\\
\Delta H_x
&=j\omega\epsilon_0
\left(\chi_\text{ee}^{yx}E_{x,\mathrm{av}}+
\chi_\text{ee}^{yy}E_{y,\mathrm{av}}\right),
\label{eq:gstc_scalar_2}\\
\Delta E_y
&=j\omega\mu_0
\left(\chi_\text{mm}^{xx}H_{x,\mathrm{av}}+
\chi_\text{mm}^{xy}H_{y,\mathrm{av}}\right),
\label{eq:gstc_scalar_3}\\
-\Delta E_x
&=j\omega\mu_0
\left(\chi_\text{mm}^{yx}H_{x,\mathrm{av}}+
\chi_\text{mm}^{yy}H_{y,\mathrm{av}}\right).
\end{align}\label{eq:GSTC_scalar}
\end{subequations}
Further assuming nongyrotropy, i.e., $\chi_\text{ee}^{xy}=\chi_\text{ee}^{yx}=\chi_\text{mm}^{xy}=\chi_\text{mm}^{yx}=0$, and considering an $x$-directed incident electric field reduce Eqs.~\eqref{eq:GSTC_scalar} to
\begin{subequations}
\begin{align}
-\Delta H_y
&=j\omega\epsilon_0\chi_{\mathrm{ee}}^{xx}E_{x,\mathrm{av}},
\\
-\Delta E_x
&=j\omega\mu_0\chi_{\mathrm{mm}}^{yy}H_{y,\mathrm{av}},
\end{align}\label{eq:nongyrotropic_GSTC}
\end{subequations}
with the dipolar fields in Eq.~\eqref{eq:incident_field_ms} being approximated in the \emph{radiative near-field zone} by their far-field expressions\footnote{One may naturally use the complete exact dipole fields~\cite{balanis2016antenna} in the synthesis of our susceptibilities, but only the dominant far-field terms---with their \emph{radiative near-field} curvature---play a significant role in this operation of WFB because the relevant distances are much larger than the reactive near field ($R<\lambda/2$,~\cite{monemi2024study}), which involves the remaining terms.}
\begin{subequations}
\begin{align}
E_x(x,y;d)=E_x(R)&=jI\ell k_0^2\exp{(-jk_0R)}/(4\pi\omega\epsilon_0R), \\
H_y(x,y;d)=H_y(R)&=jI\ell k_0 d\exp{(-jk_0R)}/(4\pi R^2), 
\end{align}
\label{eq:ff_dipole_fields}
\end{subequations}
where $R=\sqrt{x^2+y^2+d^2}$. Using the physical-source distance $d_{\mathrm{ps}}$ and the virtual-source distance $d_{\mathrm{vs}}$, we explicitly write the incident tangential fields at the metasurface as
\begin{subequations}
\begin{align}
E_x^{-}(x,y)
&=E_x(x,y;d_\text{ps}),\\
H_y^{-}(x,y)
&=H_y(x,y;d_\text{ps}),
\end{align}
\label{eq:explicit_inci_fields}
\end{subequations}
and the prescribed transmitted fields as
\begin{subequations}
\begin{align}
E_x^{+}(x,y)
&=|E_x^{-}(x,y)|\exp\left[j\angle E_x(x,y;d_\text{vs})\right],\\
H_y^{+}(x,y)
&=|H_y^{-}(x,y)|\exp\left[j\angle H_y(x,y;d_\text{vs})\right].
\end{align}
\label{eq:explicit_trans_fields}
\end{subequations}
%

\pati{Susceptibility Solution}
Equations~\eqref{eq:nongyrotropic_GSTC} provide the susceptibilities
\begin{subequations}
\begin{align}
\chi_\text{ee}^{xx}(x,y)
&=
-\frac{\Delta H_y(x,y)}
{j\omega\epsilon_0 E_{x,\mathrm{av}}(x,y)},
\label{eq:chiee_xx_phase_only}\\
\chi_\text{mm}^{yy}(x,y)
&=
-\frac{\Delta E_x(x,y)}
{j\omega\mu_0 H_{y,\mathrm{av}}(x,y)},
\label{eq:chimm_yy_phase_only}
\end{align}\label{eq:chi_solution}
\end{subequations}
where the difference and average fields are obtained by substituting Eqs.~\eqref{eq:explicit_inci_fields} and~\eqref{eq:explicit_trans_fields} into Eqs.~\eqref{eq:field_avg_jump}. The prescribed MWB transformation is thus synthesized through the electric susceptibility $\chi_{\mathrm{ee}}^{xx}$ and the magnetic susceptibility $\chi_{\mathrm{mm}}^{yy}$ of the metasurface.

\begin{figure}[h!]
\centering
\includegraphics[width=8cm]{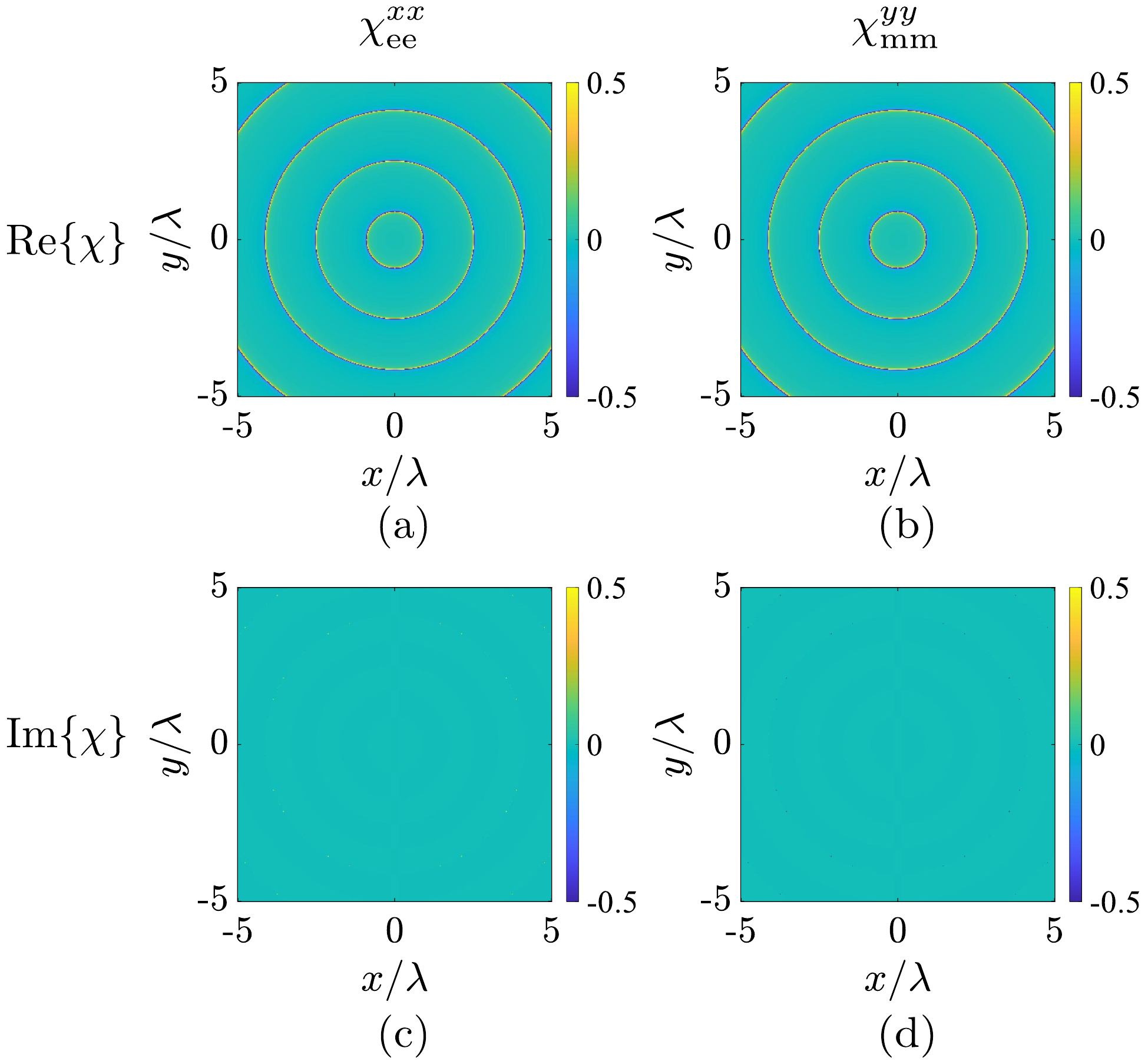}
    \caption{GSTC-synthesized WFB metasurface susceptibilities [Eqs.~\eqref{eq:chi_solution}] for $d_{\mathrm{ps}}=10\lambda$ and $\beta=8$. (a)~Real part of $\chi_\text{ee}^{xx}$. (b)~Real part of $\chi_\text{mm}^{yy}$. (c)~Imaginary part of $\chi_\text{ee}^{xx}$. (d)~Imaginary part of $\chi_\text{mm}^{yy}$.}
    \label{fig:GSTC_solution}
\end{figure}

\pati{Susceptility Distributions}
Figure~\ref{fig:GSTC_solution} shows the susceptibility distributions obtained from Eq.~\eqref{eq:chi_solution}. 
Figures~\ref{fig:GSTC_solution}(a) and~\ref{fig:GSTC_solution}(b) present the real parts of $\chi_{\mathrm{ee}}^{xx}$ and $\chi_{\mathrm{mm}}^{yy}$, respectively. The radial symmetry of the susceptibilities results from the coaxial configuration of the physical and virtual sources. The pronounced annular variations occur near locations where the incident and prescribed transmitted fields approach an antiphase condition. At these locations, the average tangential fields in the denominators of Eq.~\eqref{eq:chi_solution} is small while the corresponding field jumps remain finite, resulting in large local susceptibility values. The imaginary parts shown in Figs.~\ref{fig:GSTC_solution}(c) and~\ref{fig:GSTC_solution}(d) remain close to zero over most of the aperture, indicating that the synthesized transformation is predominantly reactive and involves negligible loss. 

\pati{Local Scattering Coefficients}

\begin{figure}[t!]
    \centering
    \begin{subfigure}[t]{0.49\linewidth}
        \includegraphics[width=\linewidth]{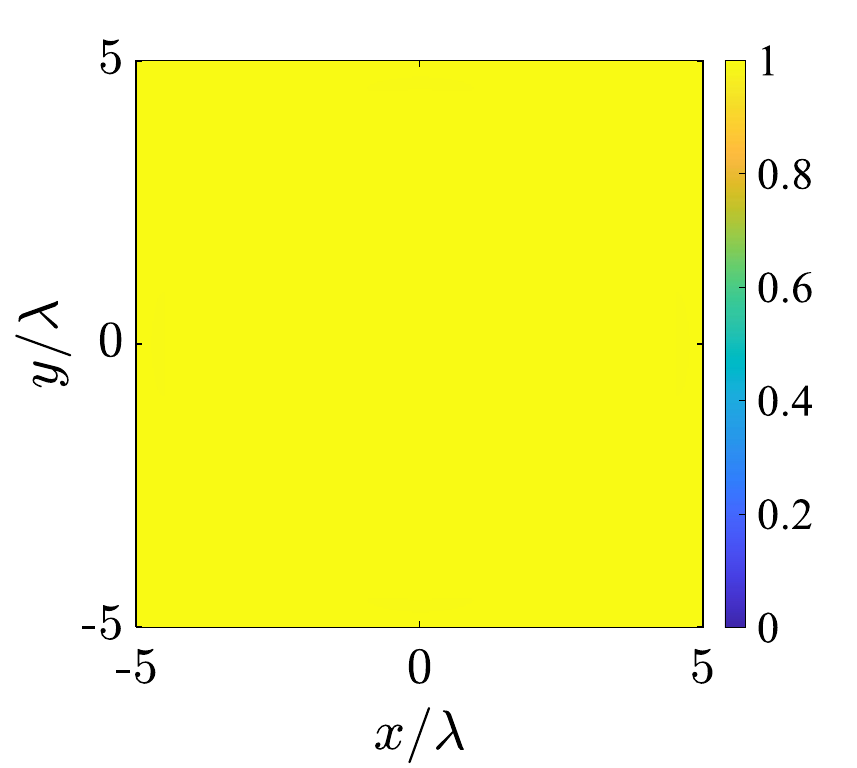}
        \captionsetup{skip=1pt}  
        \caption{}
    \end{subfigure}
    \hfill
    \begin{subfigure}[t]{0.49\linewidth}
        \includegraphics[width=\linewidth]{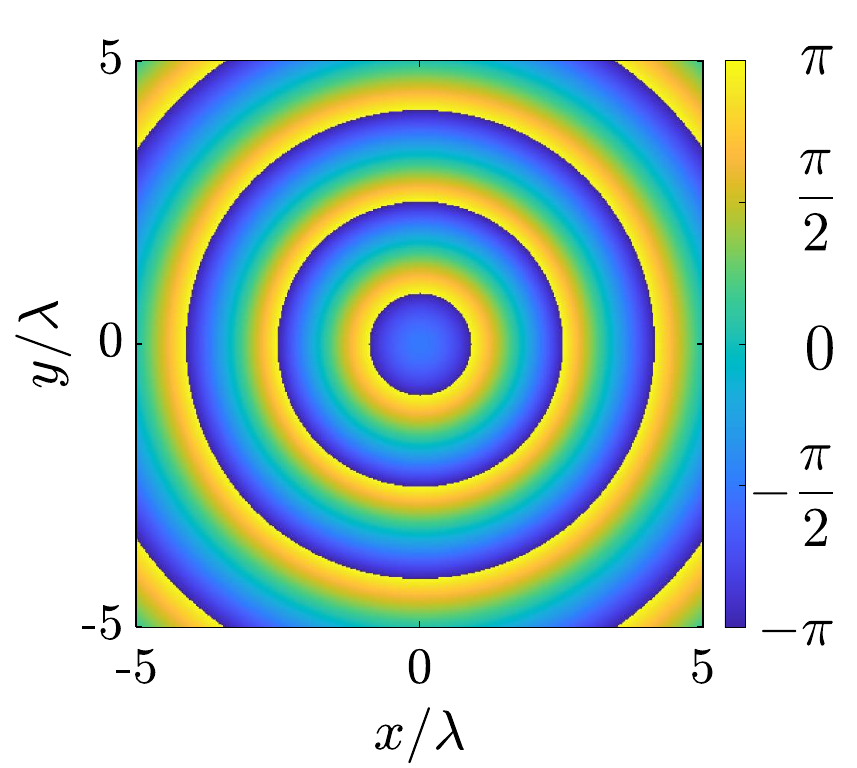}
        \captionsetup{skip=1pt}  
        \caption{}
    \end{subfigure}
    \vspace{-3mm}
    \caption{Local transmission coefficient [Eq.~\eqref{eq:local_tx_coefficient}] corresponding to the susceptibilities in Fig.\ref{fig:GSTC_solution}. (a)~Magnitude, $\lvert T\rvert$. (b)~Phase, $\angle T$.}
\label{fig:local_TR}
\end{figure}

The transmission coefficient just after the metasurface, at $z=0^+$, may be obtained by first substituting
Eqs.~\eqref{eq:explicit_inci_fields} and~\eqref{eq:explicit_trans_fields} into Eqs.~\eqref{eq:field_avg_jump}, and then substituting the resulting field averages and field jumps into Eqs.~\eqref{eq:nongyrotropic_GSTC}, which yields
\begin{equation}
T(x,y)=
\frac{4+k_0^2
\chi_\text{ee}^{xx}\chi_\text{mm}^{yy}}
{\left(2+jk_0\chi_\text{ee}^{xx}\right)
\left(2+jk_0\chi_\text{mm}^{yy}\right)}.
\label{eq:local_tx_coefficient}
\end{equation}
That is a \emph{local} transmission coefficient---a function of $x$ and $y$ at $z=0^+$, which is plotted in Fig.~\ref{fig:local_TR} for the susceptibilities in Fig.\ref{fig:GSTC_solution}
%

\section{Modeling Complete System}
\label{sec:model_withMS}
\subsection{System Configuration}
\label{sec:system_config_withMS}

\begin{figure}[!h]
\centering
\includegraphics[width=8cm]{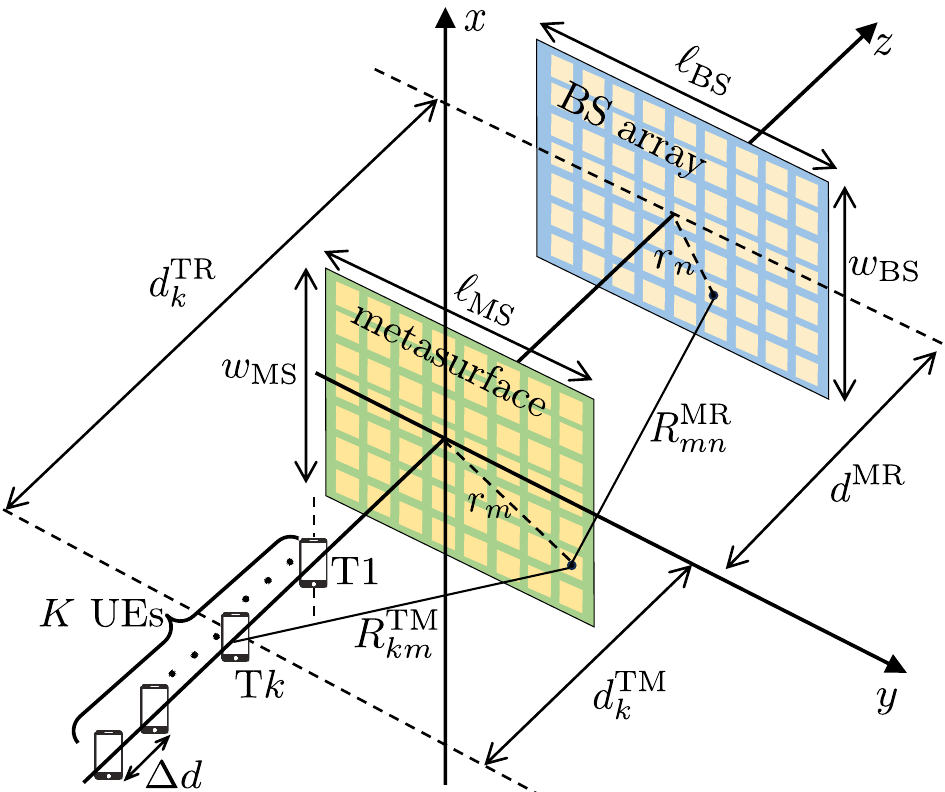}
\caption{Metasurface assisted MU-MIMO system configuration.}
\label{fig:config_withMS}
\end{figure}

\pati{System Configuration}
Figure~\ref{fig:config_withMS} shows the complete metasurface-assisted uplink MU-MIMO system, in which the $K$ UEs transmit toward the BS through a transmissive metasurface positioned between them. The metasurface lies in the $x$--$y$ plane, with its center selected as the origin of a Cartesian coordinate system. The metasurface comprises $M$ subwavelength unit cells, where unit cell $m$ is located at $\mathbf r_m^{\mathrm{MS}}=(x_m,y_m,0)$, with $m=1,\ldots,M$. The BS comprises $N$ antenna elements, where element $n$ is located at $\mathbf r_n^{\mathrm{BS}}=(x_n,y_n,d^{\mathrm{MR}})$, with $n=1,\ldots,N$. Here, $d^{\mathrm{MR}}$ denotes the axial distance between the metasurface and BS planes.
UE $k$ lies on the broadside axis at $\mathbf s_k=(0,0,-d_k^{\mathrm{TM}})$, with $k=1,\ldots,K$ and $d_k^{\mathrm{TM}}=d_1^{\mathrm{TM}}+(k-1)\Delta d$, where $d_k^{\mathrm{TM}}$ denotes the axial distance between UE~$k$ and the metasurface plane, and $\Delta d$ denotes the range spacing between adjacent UEs. The corresponding axial distance between UE~$k$ and the BS plane therefore follows as $d_k^{\mathrm{TR}}=d_k^{\mathrm{TM}}+d^{\mathrm{MR}}$.
%

\subsection{Channel Model for Common-Profile Curvature Bending}
\label{sec:channel_model_withMS}
\pati{MWB Profile Designed for Reference-user}
Following the synthesis procedure in Sec.~\ref{sec:MS_GSTC}, we select UE~$i$ as the reference UE and design the metasurface to increase the curvature of its incident wavefront by a prescribed bending ratio $\beta$. The reference UE~$i$ generates the incident field
\begin{equation}\label{eq:UEi_inci_field}
E_i^{-}(x_m,y_m)=E_x\!\left(x_m,y_m;d_i^{\mathrm{TM}}\right)\approx C_0\frac{e^{-j k_0 R^\text{TM}_{im}}}{R^\text{TM}_{im}},
\end{equation}
at metasurface element $m$, where $C_0=jk_0^2/(4\pi\omega\epsilon_0)$ under the unit-current-moment ($I\ell=1$) and $R^\text{TM}_{im}$ denotes the distance between UE~$i$ and metasurface element $m$,
\begin{equation}
R^\text{TM}_{im}=\sqrt{\left(d_i^{\mathrm{TM}}\right)^2+x_m^2+y_m^2}=\sqrt{\left(d_i^{\mathrm{TM}}\right)^2+r_m^2}.
\end{equation}
According to the transformation prescribed in Eq.~\eqref{eq:target_field_ms}, the amplitude of the incident field is preserved while the phase is associated with a virtual source at the axial distance $d_i^{\mathrm{TM}}/\beta$. The target transmitted field of UE~$i$ is therefore
\begin{equation}
E_i^{+}(x_m,y_m)
=\left|E_i^{-}(x_m,y_m)\right|\exp\left[j\angle E_x\!\left(x_m,y_m;\frac{d_i^{\mathrm{TM}}}{\beta}\right)\right].
\label{eq:UEi_trans_field}
\end{equation}
Taking the ratio of Eqs.~\eqref{eq:UEi_inci_field} and~\eqref{eq:UEi_trans_field} yields the metasurface local transmission coefficients
\begin{align}
T_{(i,\beta)}^{\mathrm{MS}}(x_m,y_m)
&=\frac{E_i^{+}(x_m,y_m)}{E_i^{-}(x_m,y_m)}\nonumber\\
&=\exp\left\{j\left[\angle E_x\!\left(x_m,y_m;\frac{d_i^{\mathrm{TM}}}{\beta}\right)\right.\right.\nonumber\\
&\left.\left.\hspace{1cm}-\angle E_x\!\left(x_m,y_m;d_i^{\mathrm{TM}}\right)\right]\right\},
\label{eq:Tms_common_profile}
\end{align}
where
\begin{subequations}
\begin{align}
\angle E_x\!\left(x_m,y_m;\frac{d_i^{\mathrm{TM}}}{\beta}\right)
&=-k_0\left(\sqrt{\left(\frac{d_i^{\mathrm{TM}}}{\beta}\right)^2+r_m^2}\right),\\
\angle E_x\!\left(x_m,y_m;d_i^{\mathrm{TM}}\right)
&=-k_0\left(\sqrt{\left(d_i^{\mathrm{TM}}\right)^2+r_m^2}\right),
\end{align}\label{eq:Tms_phase}
\end{subequations}
with $r^2_m=\sqrt{x^2_m+y^2_m}$.
Substituting Eq.~\eqref{eq:Tms_phase} into Eq.~\eqref{eq:Tms_common_profile} gives
\begin{align}
T_{(i,\beta)}^{\mathrm{MS}}(x_m,y_m)=
&\exp\!\left\{-jk_0\left[\sqrt{\left(\frac{d_i^{\mathrm{TM}}}{\beta}\right)^2+r_m^2}\right.\right.\nonumber\\
&\qquad\left.\left.-\sqrt{\left(d_i^{\mathrm{TM}}\right)^2+r_m^2}\right]\right\}.
\label{eq:Tms_expand}
\end{align}
We will use Eq.~\eqref{eq:Tms_expand} as a common transmission profile and apply it to the incident fields of all UEs.

\pati{UE-MS Propagation}
We now consider an arbitrary UE~$k$ and derive its channel to the BS array through the metasurface. UE $k$ generates the incident field
\begin{equation}\label{eq:UEk_inci_field}
E_k^{-}(x_m,y_m)=E_x\!\left(x_m,y_m;d_k^{\mathrm{TM}}\right)\approx C_0\frac{e^{-j k_0 R^\text{TM}_{km}}}{R^\text{TM}_{km}},
\end{equation}
at metasurface element $m$, where $R^\text{TM}_{km}$ denotes the distance between UE $k$ and metasurface element $m$,
\begin{equation}
R^\text{TM}_{km}=\sqrt{\left(d_k^{\mathrm{TM}}\right)^2+x_m^2+y_m^2}=\sqrt{\left(d_k^{\mathrm{TM}}\right)^2+r_m^2}.
\end{equation}
%

\pati{MWB Transformation}
The common metasurface profile transforms the incident field of UE~$k$ into the transmitted field,
\begin{equation}
E_k^{+}(x_m,y_m)
=E_k^{-}(x_m,y_m)T_{(i,\beta)}^{\mathrm{MS}}(x_m,y_m),
\label{eq:field_after_ms_k}
\end{equation}
where $E_k^{-}(x_m,y_m)$ and $T_{(i,\beta)}^{\mathrm{MS}}(x_m,y_m)$ are given by Eqs.~\eqref{eq:UEk_inci_field} and~\eqref{eq:Tms_expand}, respectively.

\pati{MS-BS Propagation and Channel Assembly}
The transmitted field $E_k^{+}(x_m,y_m)$ subsequently propagates from the metasurface to the BS array. We model this stage using a Rayleigh--Sommerfeld-like scalar diffraction formulation~\cite{born2013principles}, which gives the channel coefficient between UE~$k$ and BS element $n$ as
\begin{equation}
h_{kn}=\frac{1}{j\lambda}\iint_{\mathcal{A}_{\mathrm{MS}}}E_k^{+}(x_m,y_m)\frac{e^{-jk_0R_{mn}^{\mathrm{MR}}}}{R_{mn}^{\mathrm{MR}}}\frac{d^{\mathrm{MR}}}{R_{mn}^{\mathrm{MR}}}\,\mathrm{d}x_m\,\mathrm{d}y_m,
\label{eq:complete_channel_continuous}
\end{equation}
where the factor $d^{\mathrm{MR}}/R$ is the Rayleigh--Sommerfeld obliquity factor, which weights the contribution of each aperture point according to its direction with respect to BS element $n$. We assume that each metasurface unit cell has area $\Delta S$ and discretize Eq.~\eqref{eq:complete_channel_continuous} over the $M$ elements. Substituting this discretization, $dx_mdy_m=\Delta S$  and Eq.~\eqref{eq:field_after_ms_k} into Eq.~\eqref{eq:complete_channel_continuous} gives
\begin{equation}
h_{kn}
\approx
\frac{\Delta S}{j\lambda}\sum_{m=1}^{M}E_k^{-}(x_m,y_m)T_{(i,\beta)}^{\mathrm{MS}}(x_m,y_m)
\frac{e^{-jk_0R_{mn}^{\mathrm{MR}}}}{R_{mn}^{\mathrm{MR}}}
\frac{d^{\mathrm{MR}}}{R_{mn}^{\mathrm{MR}}}.
\label{eq:complete_channel_dis}
\end{equation}
Substituting then Eq.~\eqref{eq:UEk_inci_field} and Eq.~\eqref{eq:Tms_expand} into Eq.~\eqref{eq:complete_channel_dis} yields the explicit expression of the channel as
\begin{align}
h_{kn}(\beta)
&\approx\frac{\Delta S}{j\lambda}\sum_{m=1}^{M}\frac{C_0}{R^\text{TM}_{km}}\frac{d^{\mathrm{MR}}}{\left(R_{mn}^{\mathrm{MR}}\right)^2}\nonumber\\
&\exp\left\{-jk_0\left[R^\text{TM}_{km}+R_{mn}^{\mathrm{MR}} \right.\right.\nonumber\\
&\qquad+\left.\left.\left(\sqrt{\left(\frac{d_i^{\mathrm{TM}}}{\beta}\right)^2+r_m^2}-\sqrt{\left(d_i^{\mathrm{TM}}\right)^2+r_m^2}\right)\right]\right\},
\label{eq:h_kn_beta_explicit}
\end{align}
where $R_{mn}^{\mathrm{MR}}$ is the distance between metasurface element $m$ and BS element $n$,
\begin{align}
R_{mn}^{\mathrm{MR}}&=\sqrt{\left(d^{\mathrm{MR}}\right)^2+(x_n-x_m)^2+(y_n-y_m)^2}\nonumber\\ &=\sqrt{\left(d^{\mathrm{MR}}\right)^2+r^2_{mn}}.
\label{eq:R_MB_continuous}
\end{align}
For compactness, we shall later collect the amplitude factor in the expression~\eqref{eq:h_kn_beta_explicit} as
\begin{equation}
A_{kmn}
=\frac{\Delta S C_0 d^{\mathrm{MR}}}{j\lambda R^\text{TM}_{km}\left(R_{mn}^{\mathrm{MR}}\right)^2}.
\label{eq:common_amplitude_factor}
\end{equation}

Particularly, the channel coefficient of the reference UE ($k=i$) takes the simpler form
\begin{align}
h_{in}(\beta)=A_{imn}\exp\left\{-jk_0\left[R_{mn}^{\mathrm{MR}}+\sqrt{\left(\frac{d_i^{\mathrm{TM}}}{\beta}\right)^2+r_m^2} \right]\right\},
\label{eq:h_in_beta_explicit}
\end{align}

%
\pati{Channel Assembly under Phase Approximation}
To separate the phase contributions of the different propagation stages and identify their dependence on the bending ratio, we approximate the phase terms in Eq.~\eqref{eq:h_kn_beta_explicit} as
\begin{subequations}
\begin{align}
&k_0R^\text{TM}_{km}=k_0\left(\sqrt{\left(d_k^{\mathrm{TM}}\right)^2+r_m^2}\right)\approx k_0\left(d_k^{\mathrm{TM}}+\frac{r_m^2}{2d_k^{\mathrm{TM}}}\right),\\
&k_0R_{mn}^{\mathrm{MR}}=k_0\left(\sqrt{\left(d^{\mathrm{MR}}\right)^2+r_{mn}^2}\right)\approx k_0\left(d^{\mathrm{MR}}+\frac{r_{mn}^2}{2d^{\mathrm{MR}}}\right),\\
&k_0\left(\sqrt{\left(\frac{d_i^{\mathrm{TM}}}{\beta}\right)^2+r_m^2}\right)
\approx k_0\left(\frac{d_i^{\mathrm{TM}}}{\beta}+\frac{r_m^2}{2d_i^{\mathrm{TM}}/\beta}\right),\\
&k_0\left(\sqrt{\left(d_i^{\mathrm{TM}}\right)^2+r_m^2}\right)
\approx k_0\left(d_i^{\mathrm{TM}}+\frac{r_m^2}{2d_i^{\mathrm{TM}}}\right).
\end{align}\label{eq:TMR_phase_approx}
\end{subequations}
Substituting Eqs.~\eqref{eq:TMR_phase_approx} and~\eqref{eq:common_amplitude_factor} into Eq.~\eqref{eq:h_kn_beta_explicit} gives then phase-approximated channel of UE $k$ as
\begin{align}
h_{kn}(\beta)
&\approx \sum_{m=1}^{M}A_{kmn}\exp\!\left\{-jk_0\left[\left(d_k^{\mathrm{TM}}+\frac{r_m^2}{2d_k^{\mathrm{TM}}}\right)\right.\right.\nonumber\\
&\qquad+\left(\frac{d_i^{\mathrm{TM}}}{\beta}+\frac{\beta r_m^2}{2d_i^{\mathrm{TM}}}\right)-\left(d_i^{\mathrm{TM}}+\frac{r_m^2}{2d_i^{\mathrm{TM}}}\right)\nonumber\\
&\left.\left.\qquad+\left(d^{\mathrm{MR}}+\frac{r_{mn}^2}{2d^{\mathrm{MR}}}\right)\right]\right\}.
\label{eq:h_kn_beta_approx}
\end{align}
The first term inside the square brackets represents the propagation phase from UE~$k$ to metasurface element $m$. The difference between the second and third terms represents the common curvature-bending phase imposed by the reference UE  $i$. The final term represents the propagation phase from metasurface element $m$ to BS element $n$.
Particularly, the channel coefficient of the reference UE ($k=i$) takes the simpler form
\begin{align}
h_{in}(\beta)
&\approx \sum_{m=1}^{M}A_{imn}\exp\!\left\{-jk_0\left[\left(\frac{d_i^{\mathrm{TM}}}{\beta}+\frac{\beta r_m^2}{2d_i^{\mathrm{TM}}}\right)\right.\right.\nonumber\\
&\left.\left.\hspace{3cm}+\left(d^{\mathrm{MR}}+\frac{r_{mn}^2}{2d^{\mathrm{MR}}}\right)\right]\right\}.
\label{eq:common_hni_explicit}
\end{align}

\section{Demonstration}
\label{sec:performance_withMS}
\pati{Motivation}
This section uses the complete channel model in Sec.~\ref{sec:model_withMS} to quantify the multiplexing gains achieved with MWB. The evaluation first considers a practical common-profile scheme, where a MWB profile is designed for one reference UE and then applied to all UEs. We then compare this practical scheme with a user-specific MWB benchmark, where each UE is assigned an ideal curvature-bending profile. The benchmark clarifies the spatial multiplexing gain attainable when distinct curvature signatures can be assigned to different users.

\subsection{Performance under Common-Profile Curvature Bending}
\begin{figure}[!h]
    \centering
    \begin{subfigure}[t]{0.49\linewidth}
        \includegraphics[width=\linewidth]{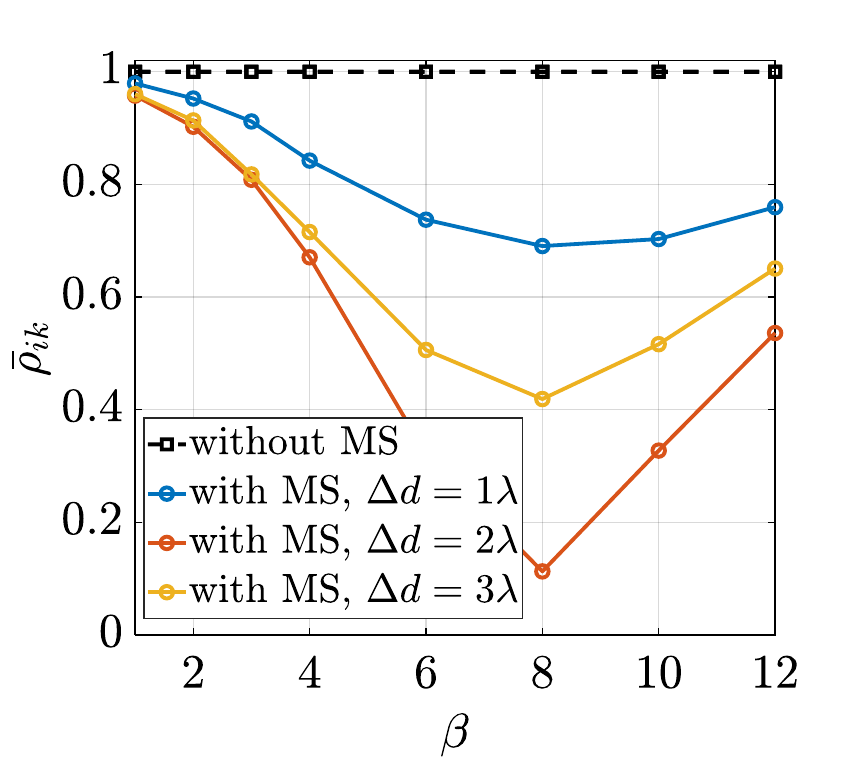}
        \captionsetup{skip=1pt}
        \caption{}
    \end{subfigure}
    \hfill
    \begin{subfigure}[t]{0.49\linewidth}
        \includegraphics[width=\linewidth]{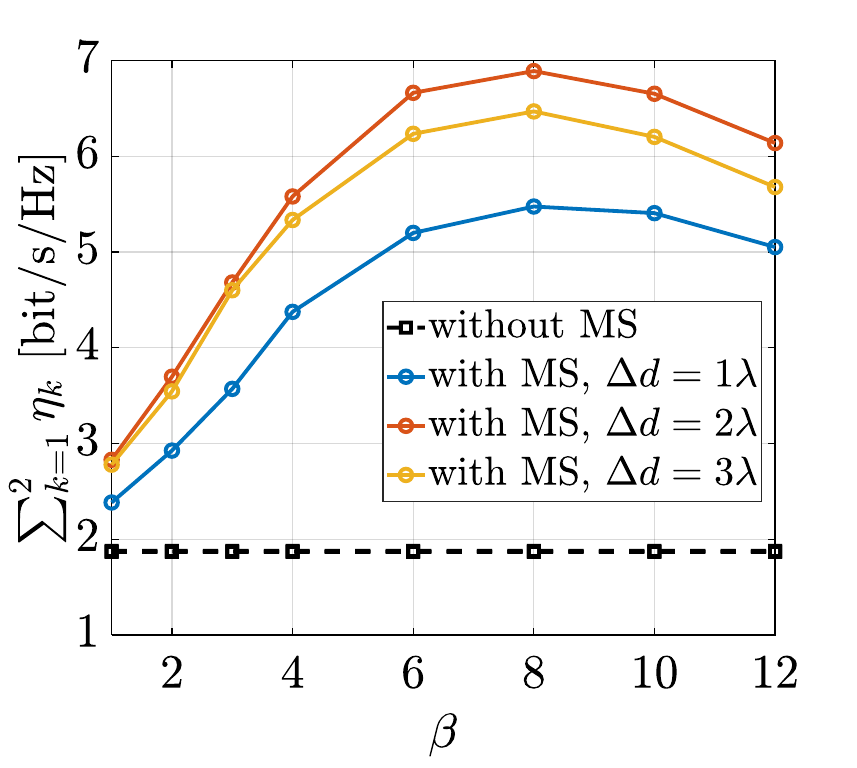}
        \captionsetup{skip=1pt}
        \caption{}
    \end{subfigure}
\caption{Two-user MIMO performance for common-profile MWB as a function of the bending ratio $\beta$, obtained for the system configuration shown in Fig.~\ref{fig:config_withMS}. 
(a)~Normalized channel correlation, $\bar{\rho}_{ik}$ [Eq.~\eqref{eq:normalized_rho}]. 
(b)~Spectral efficiency sum, $\eta_i+\eta_k$ [Eq.~\eqref{eq:SE}], with the interference-free reference SNR of $\gamma_i=\gamma_k=10~\mathrm{dB}$. 
The metasurface profile is designed for the reference UE $i$ and applied to both UEs as a common profile. The simulation parameters are $f_0=10~\mathrm{GHz}$, $\ell_{\mathrm{MS}}=w_{\mathrm{MS}}=10\lambda$, with a unit-cell spacing of $\lambda/5$,  $\ell_{\mathrm{BS}}=w_{\mathrm{BS}}=10\lambda$, with an antenna-element spacing of $\lambda/2$, $d^{\mathrm{TM}}_i=5\lambda$, and $d^{\mathrm{MR}}=50\lambda$, $d^{\mathrm{TR}}_i=d^{\mathrm{TM}}_i+d^{\mathrm{MR}}=55\lambda$.}
\label{fig:two_user_common_profile}
\end{figure}

\pati{$\beta$-Dependent Performance for Two-User MIMO}
Consider the reference UE~$i$ and an arbitrary UE~$k$. Figure~\ref{fig:two_user_common_profile} plots the corresponding two-user performance as a function of the bending ratio. Figure~\ref{fig:two_user_common_profile}(a) presents the normalized channel correlation in Eq.~\eqref{eq:normalized_rho} and Fig.~\ref{fig:two_user_common_profile}(b) presents the spectral efficiency sum using Eq.~\eqref{eq:SE}, with the channel coefficients given by Eqs.~\eqref{eq:h_kn_beta_explicit} and~\eqref{eq:h_in_beta_explicit}. Figure~\ref{fig:two_user_common_profile}(a) shows that, without the metasurface, the normalized channel correlation remains high. This occurs because in the considered configuration, the reference UE $i$ lies at an axial distance of $d_i^{\mathrm{TR}}=55\lambda$ from the BS, which yields only weak spherical-wave curvature across the BS aperture and therefore provides insufficient spatial variation for distinguishing channel vectors. After applying the common-profile WFB metasurface, $\bar{\rho}_{ik}$ initially decreases as $\beta$ increases, reaches a minimum at an intermediate bending ratio, and then increases at larger values of $\beta$. The dependence on the range separation $\Delta d$ also remains nonmonotonic over the considered bending-ratio range. The strongest decorrelation occurs for $\Delta d=2\lambda$, for which the channel correlation decreases to approximately $0.1$ at $\beta=8$.
The spectral efficiency sum in Fig.~\ref{fig:two_user_common_profile}(b) exhibits the inverse trend. It increases as the channel correlation decreases, reaches its maximum close to the correlation minimum, and then decreases when the correlation rises again.
%

\pati{Explanation of the Nonmonotonic Dependence on $\beta$}
To explain the nonmonotonic dependence on $\beta$ observed in Fig.~\ref{fig:two_user_common_profile}, we explicitly derive the elementwise cross-channel product $h_{in}^{*}h_{kn}$ using Eq.~\eqref{eq:common_hni_explicit} and Eq.~\eqref{eq:h_kn_beta_approx}. This product takes the form
\begin{align}
h_{in}^{*}h_{kn}
&\approx e^{-jk_0\left(d_k^{\mathrm{TM}}-d_i^{\mathrm{TM}}\right)}
\sum_{m=1}^{M}\sum_{q=1}^{M}
A_{iqn}^{*}A_{kmn}\nonumber\\
&\quad\exp\!\left\{\!\!-jk_0\left[\left(\frac{1}{2d_k^{\mathrm{TM}}}-\frac{1}{2d_i^{\mathrm{TM}}}\right)r_m^2 \right.\right.\nonumber\\
&\qquad+\frac{\beta}{2d_i^{\mathrm{TM}}}\left(r_m^2-r_q^2\right)\nonumber\\[-1mm]
&\left.\left.\qquad+\frac{1}{2d^{\mathrm{MR}}}\left(r_{mn}^2-r_{qn}^2\right)\right]\right\}.
\label{eq:common_phase_product}
\end{align}
The exponent in Eq.~\eqref{eq:common_phase_product} contains three aperture-dependent phase contributions.
The first contribution results from the range mismatch between the reference and non-reference UEs. The second contribution is directly controlled by the bending ratio $\beta$. The third contribution arises from propagation between the metasurface and the BS array.
The numerator of the normalized correlation in Eq.~\eqref{eq:normalized_rho} is obtained by summing Eq.~\eqref{eq:common_phase_product} over the BS-element index $n$. It is therefore a finite, geometry-dependent phasor sum over the metasurface indices $m$ and $q$ and the BS index $n$. Its value depends jointly on $\beta$, $\Delta d$, $d^{\mathrm{MR}}$, and the metasurface and BS aperture dimensions. Consequently, neither the bending ratio nor the range separation alone determines the resulting correlation.
As $\beta$ increases from unity, the second phase term in Eq.~\eqref{eq:common_phase_product} produces a larger phase dispersion.
The corresponding phasors then add less coherently in Eq.~\eqref{eq:normalized_rho}, causing the channel correlation to decrease.
When $\beta$ becomes very large, however, the bending-related phase factor wraps rapidly across the finite metasurface aperture and interacts with the fixed range-mismatch and MS--BS propagation factors. The phasor sum can then become partially coherent again, producing the increase in $\bar{\rho}_{ik}$ observed at large $\beta$ in Fig.~\ref{fig:two_user_common_profile}(a). The same expression also explains the nonmonotonic dependence on $\Delta d$: increasing $\Delta d$ enlarges the first phase term in Eq.~\eqref{eq:common_phase_product} and strengthens decorrelation over a useful range, but excessive mismatch can again produce phase wrapping rather than monotonic improvement.

\begin{figure*}[t!]
    \centering

    \begin{subfigure}[t]{0.28\textwidth}
        \centering
        \includegraphics[width=\linewidth]
        {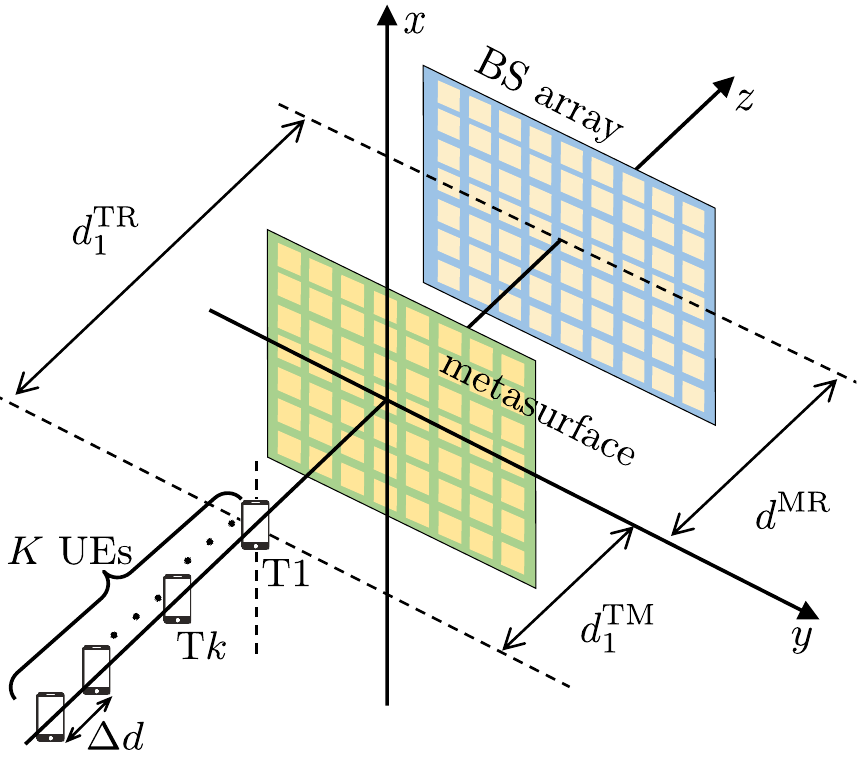}
        \captionsetup{skip=1pt}
        \caption{}
        \label{fig:same_direction_config}
    \end{subfigure}
    \hspace{0.02\textwidth}
    \begin{subfigure}[t]{0.26\textwidth}
        \centering
        \includegraphics[width=\linewidth]
        {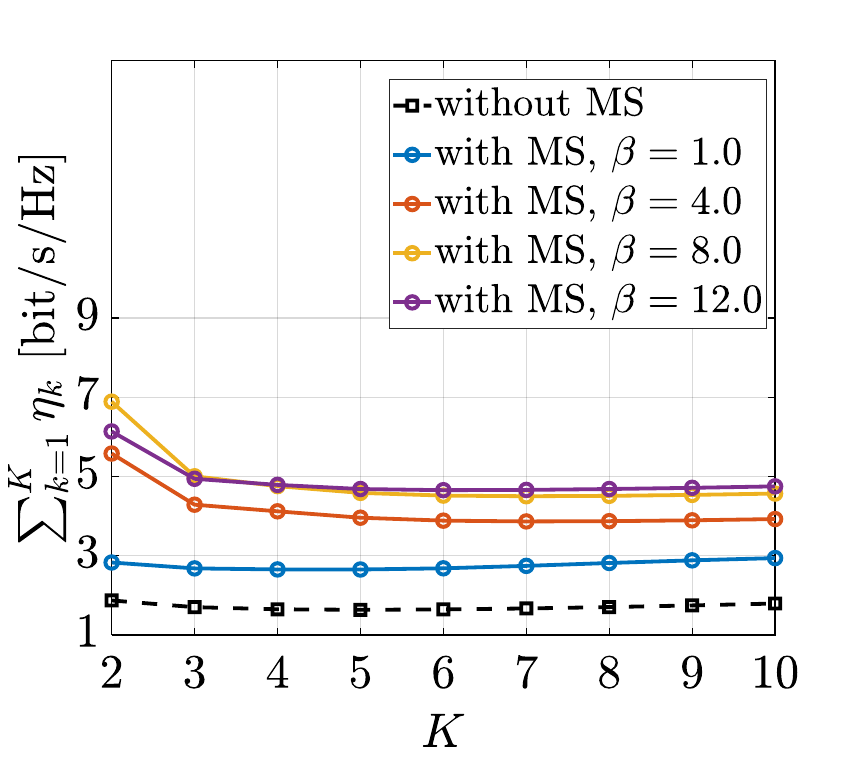}
        \captionsetup{skip=1pt}
        \caption{}
        \label{fig:same_direction_SE}
    \end{subfigure}
    \hspace{0.012\textwidth}
    \begin{subfigure}[t]{0.26\textwidth}
        \centering
        \includegraphics[width=\linewidth]
        {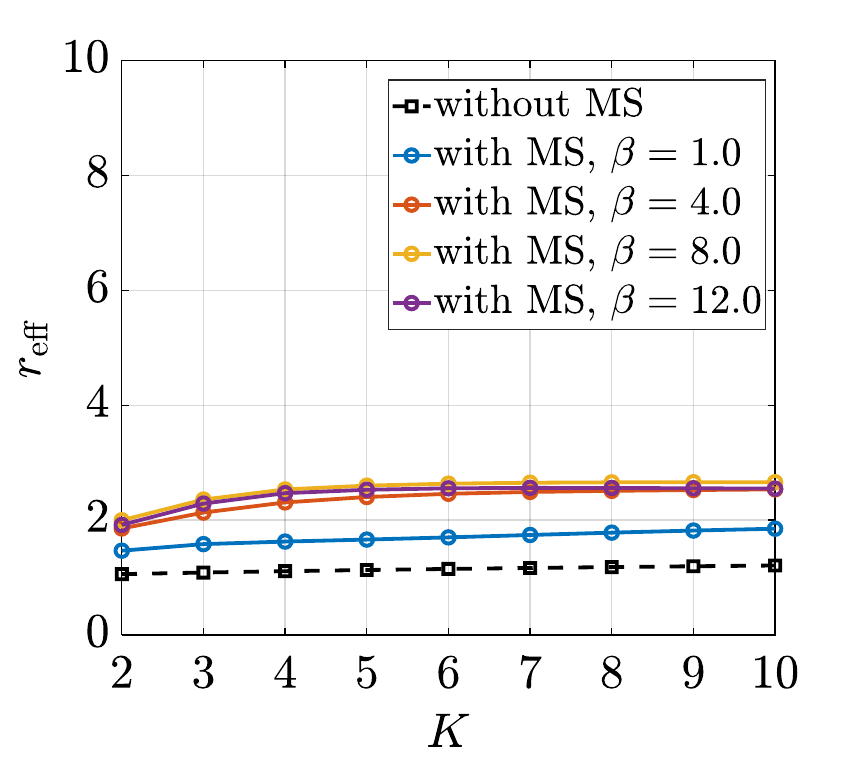}
        \captionsetup{skip=1pt}
        \caption{}
        \label{fig:same_direction_rank}
    \end{subfigure}

    \vspace{1mm}
    \begin{subfigure}[t]{0.28\textwidth}
        \centering
        \includegraphics[width=\linewidth]
        {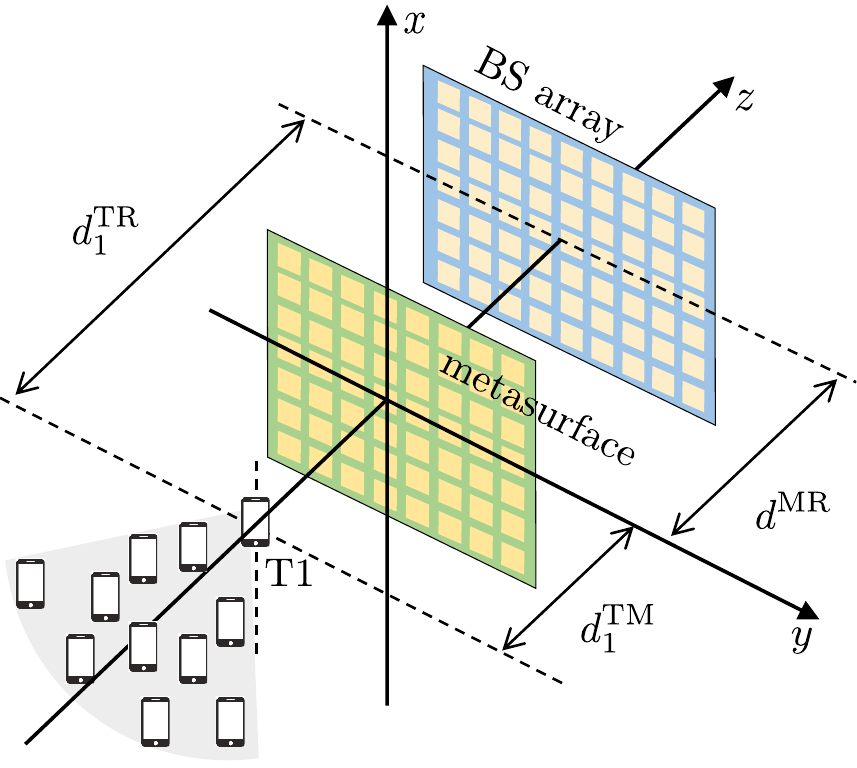}
        \captionsetup{skip=1pt}
        \caption{}
        \label{fig:random_sector_config}
    \end{subfigure}
   \hspace{0.02\textwidth}
    \begin{subfigure}[t]{0.26\textwidth}
        \centering
        \includegraphics[width=\linewidth]
        {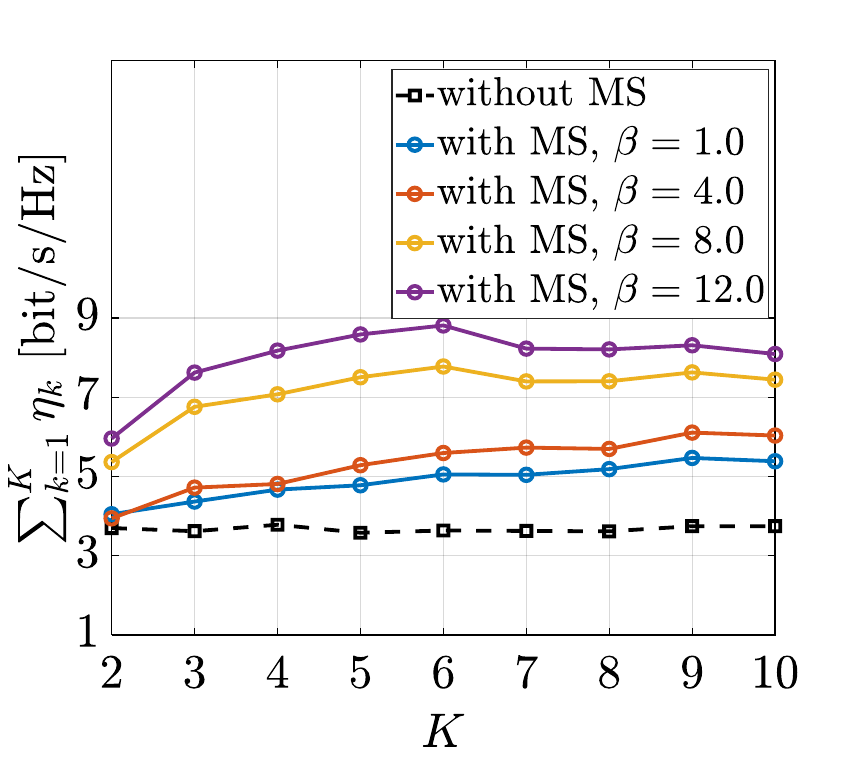}
        \captionsetup{skip=1pt}
        \caption{}
        \label{fig:random_sector_SE}
    \end{subfigure}
   \hspace{0.012\textwidth}
    \begin{subfigure}[t]{0.26\textwidth}
        \centering
        \includegraphics[width=\linewidth]
        {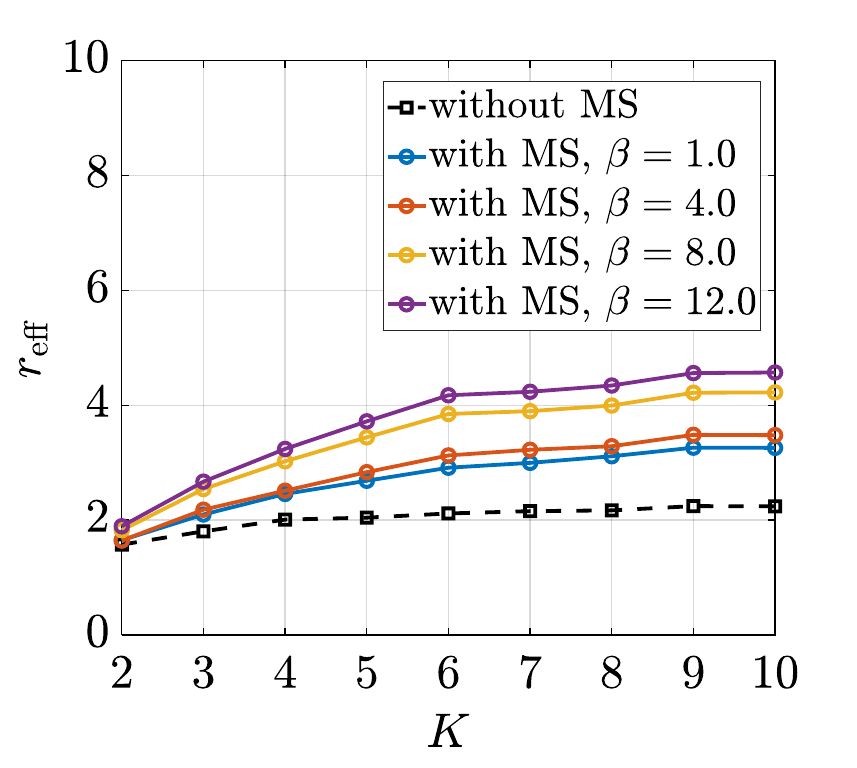}
        \captionsetup{skip=1pt}
        \caption{}
        \label{fig:random_sector_rank}
    \end{subfigure}

    \vspace{-2mm}

    \caption{Multi-user performance for the common-profile MWB scheme for different bending ratios $\beta$. First row: $K$ same-direction UEs with adjacent range spacing $\Delta d=2\lambda$: (a)~system configuration,
    (b)~spectral efficiency sum, and (c)~effective channel rank. Second row: UEs randomly distributed in the specified
    range-angle sector: (d)~system configuration, (e)~spectral efficiency sum, and (f)~effective channel rank. The UE-to-MS-center distance is independently and uniformly distributed over $d_i^{\mathrm{TM}}\in[5\lambda,12\lambda]$, while the angular position is independently and uniformly distributed over $\theta_k\in[-15^{\circ},15^{\circ}]$, corresponding to a sector centered around the broadside direction with a total angular span of $30^{\circ}$. The remaining system parameters are the same as those in Fig.~\ref{fig:two_user_common_profile}.}
    \label{fig:multi_user_common_profile}
\end{figure*}

\pati{Performance for MU-MIMO}
The common-profile MWB scheme is next evaluated for arbitrary $K~(K\ge2)$. The closest UE is selected as the reference UE, and the same metasurface from Eq.~\eqref{eq:Tms_expand} is applied to all users. For each UE, the channel vector is assembled from Eq.~\eqref{eq:h_kn_beta_explicit}. Figure~\ref{fig:multi_user_common_profile} reports the resulting MU-MIMO performance, where the spectral efficiency sum follows Eq.~\eqref{eq:SE} and the effective channel rank follows Eq.~\eqref{eq:effective_rank}. The first row in the figure involves $K$ UEs located along the same broadside direction, with an adjacent range spacing $\Delta d =2\lambda$. 
Without the metasurface, the spectral efficiency sum and effective rank remain low because the UE channels are nearly parallel. The common-profile MWB scheme provides a clear improvement over the no-MS case. Increasing the bending ratio further enhances both metrics, indicating that the curvature-bending metasurfaces creates additional spatial diversity even when the users are angularly aligned.
The second row pertains to UEs that are randomly distributed within a range-angle sector. Unlike the same-direction arrangement, these UEs possess natural diversity in both range and angle, so the no-MS baseline already benefits from angular separation. The common-profile MWB scheme further enhances the multiplexing by introducing curvature-induced diversity in addition to the existing angular diversity.
In both rows, the incremental benefit of MWB tends to saturate as the number of users $K$ increases because a single common metasurface cannot generate a new independent spatial mode for every additional UE. Nevertheless, for any fixed $K$, the bending-induced phase diversity reduces the overall channel similarity and enhances the multiplexing capability relative to the no-MS case.
%

\subsection{Performance under User-Specific Curvature Bending}
\pati{Definition of User-Specific Curvature Bending}
The common-profile MWB scheme uses the same metasurfaces for all UEs. This is the simplest practical configuration, but it limits in the maximum achievable multiplexing gain. To quantify the additional gain available from independent wavefront-curvature control, we here introduce an ideal user-specific MWB benchmark. For UE $k$, the transformation profile is obtained by replacing the reference distance in Eq.~\eqref{eq:Tms_expand} with $d_k^{\mathrm{TM}}$ and by assigning a user-specific bending ratio $\beta_k$, which yields
\begin{align}
T_{(k,\beta_k)}^{\mathrm{US}}(x_m,y_m)
&=\exp\!\left\{j\left[\angle E_x\!\left(x_m,y_m;\frac{d_k^{\mathrm{TM}}}{\beta_k}\right) \right.\right.\nonumber\\
&\qquad \left.\left.-\angle E_x\!\left(x_m,y_m;d_k^{\mathrm{TM}}\right)
\right]\right\}.
\label{eq:user_specific_mask}
\end{align}
The corresponding channel vector, denoted by $\mathbf h_k^{\mathrm{US}}(\beta_k)$, is evaluated using the complete UE--metasurface--BS channel model in Eq.~\eqref{eq:complete_channel_dis}, with the bending profile designed specifically for UE $k$. For UEs $i$ and $k$, the normalized channel correlation is
\begin{equation}
\bar{\rho}_{ik}(\beta_i,\beta_k)
=
\frac{
\left|
\left[
\mathbf h_i^{\mathrm{US}}(\beta_i)
\right]^{\mathrm H}
\mathbf h_k^{\mathrm{US}}(\beta_k)
\right|
}{
\left\|
\mathbf h_i^{\mathrm{US}}(\beta_i)
\right\|
\left\|
\mathbf h_k^{\mathrm{US}}(\beta_k)
\right\|
}.
\label{eq:user_specific_channel_correlation}
\end{equation}
The bending-ratio assignment proceeds sequentially. The first UE is assigned a specific bending ratio, $\beta_1$. When UE $k~(k\geq2)$ is added, the previously assigned bending ratios $\beta_1,\ldots,\beta_{k-1}$ remain fixed. Eq.~\eqref{eq:complete_channel_dis} generates the channel of UE $k$ from each available candidate $\beta\in\mathcal B$, and Eq.~\eqref{eq:user_specific_channel_correlation} evaluates its correlations with all previously assigned $k-1$ UE channels. The selected bending ratio for UE $k$ minimizes the largest pairwise correlation:
\begin{equation}
\beta_k
=
\underset{
\beta\in\mathcal B\setminus
\{\beta_1,\ldots,\beta_{k-1}\}
}{
\arg\min
}
\;
\max_{1< i<k}
\bar{\rho}_{ik}(\beta_i,\beta).
\label{eq:user_specific_beta_selection}
\end{equation}
Thus, each newly introduced UE is assigned the bending ratio that minimizes its largest pairwise channel correlation with the preceding $k-1$ UEs. The resulting bending-ratio sets for $K=2,\ldots,10$ are listed in Appendix~\ref{app:selected_user_specific_beta}. Because the benchmark assigns a different phase profile to each UE, it represents an ideal upper-bound reference.

\begin{figure}[h!]
    \centering
    \begin{subfigure}[t]{0.49\linewidth}
        \includegraphics[width=\linewidth]{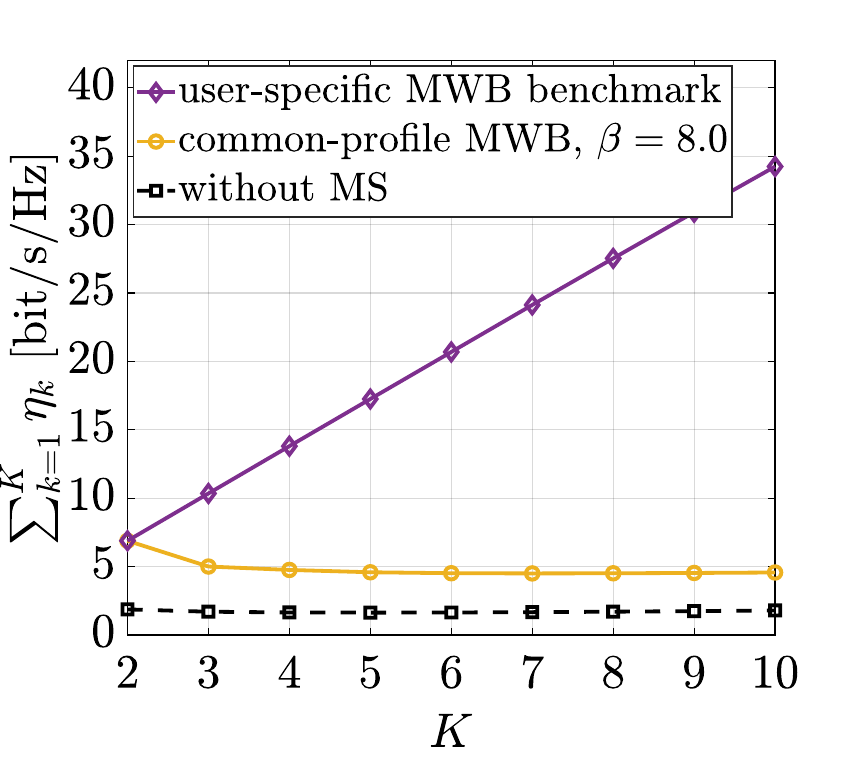}
        \captionsetup{skip=1pt}  
        \caption{}
    \end{subfigure}
    \hfill
    \begin{subfigure}[t]{0.49\linewidth}
        \includegraphics[width=\linewidth]{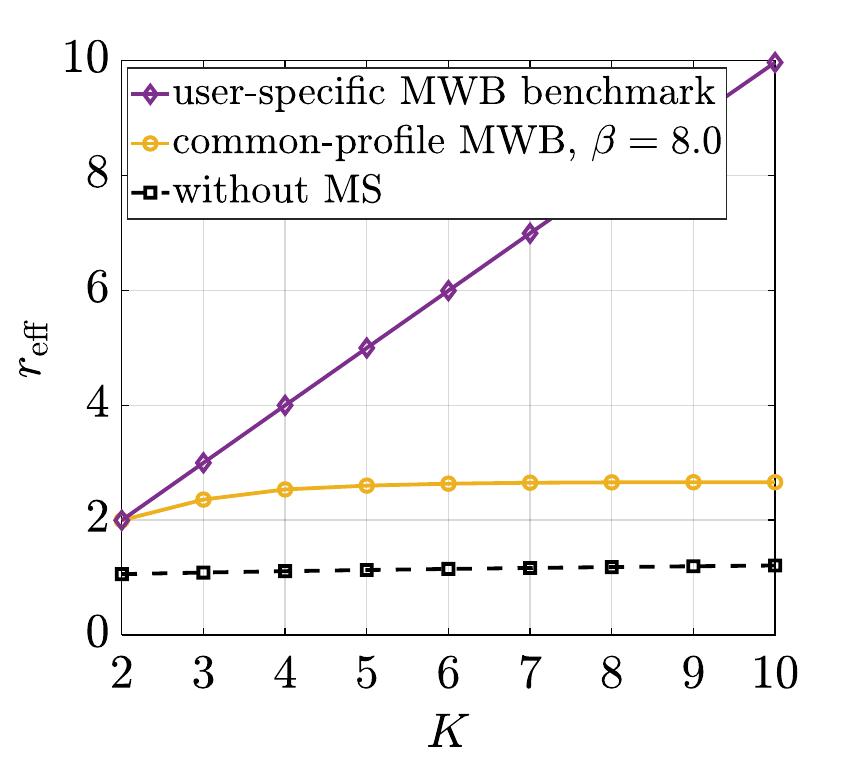}
        \captionsetup{skip=1pt}  
        \caption{}
    \end{subfigure}
    \vspace{1mm} 
    \begin{subfigure}[t]{0.49\linewidth}
        \includegraphics[width=\linewidth]{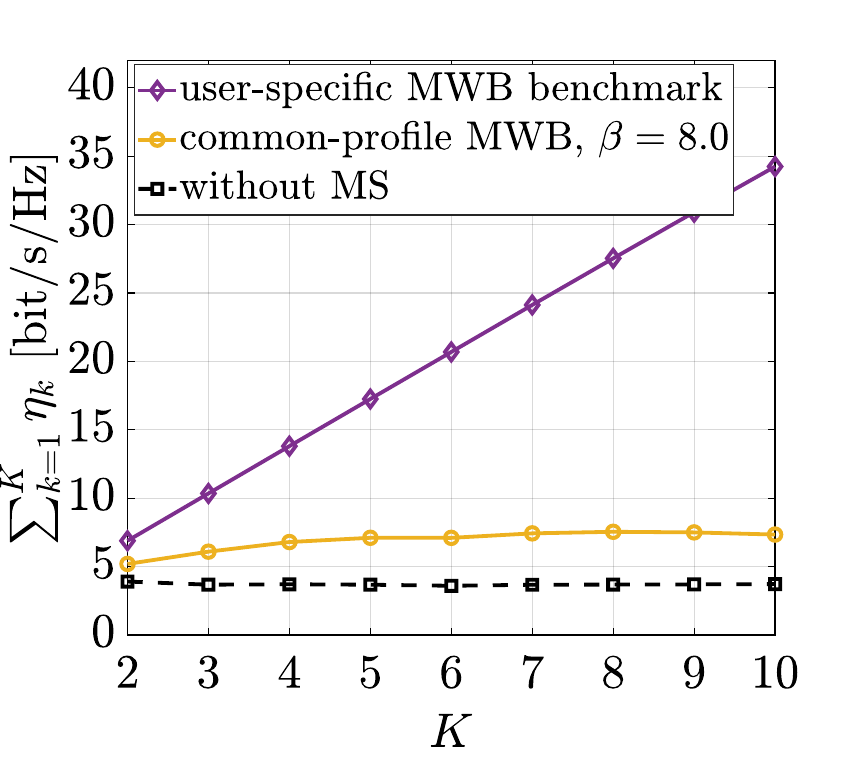}
        \captionsetup{skip=1pt}  
        \caption{}
    \end{subfigure}
    \hfill
    \begin{subfigure}[t]{0.49\linewidth}
        \includegraphics[width=\linewidth]{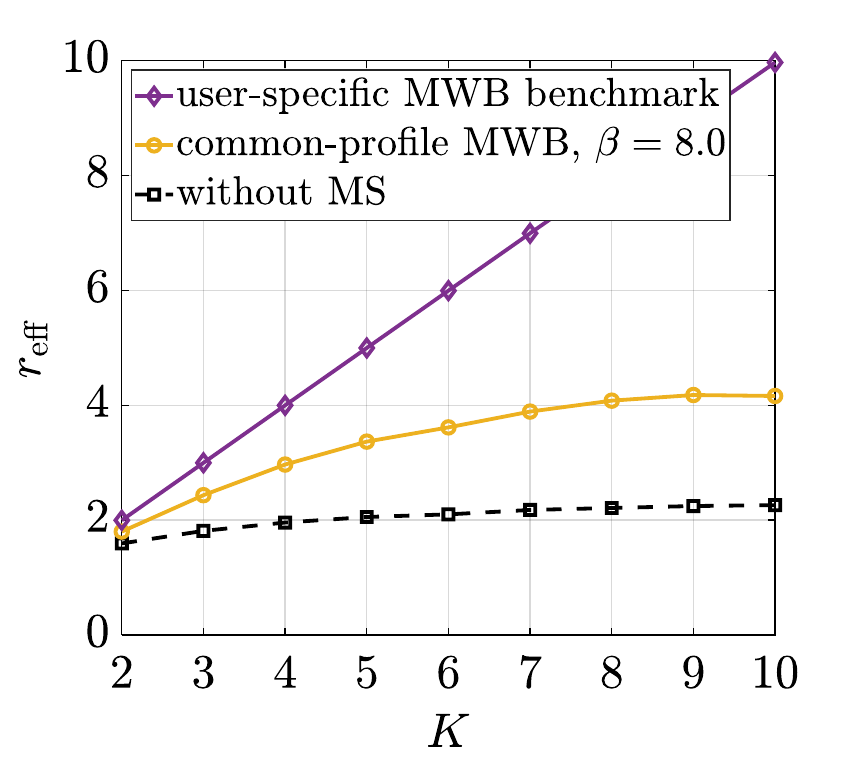}
        \captionsetup{skip=1pt}  
        \caption{}
    \end{subfigure}
    \vspace{-3mm}
    \caption{Comparison of the multi-user performance achieved by the user-specific MWB benchmark and the common-profile MWB scheme with $\beta=8$. First row: same-direction UE configuration (in Fig.~\ref{fig:multi_user_common_profile}(a)): (a)~spectral efficiency sum and (b)~effective channel rank. Second row: randomly distributed UE configuration (Fig.~\ref{fig:multi_user_common_profile}(d)): (c)~spectral efficiency sum and (d) effective channel rank.}
\label{fig:benchmark_performance}
\end{figure}

\pati{Performance Upper Bound}
Figure~\ref{fig:benchmark_performance} compares the no-MS case, the common-profile MWB scheme with $\beta=8$, and the user-specific MWB benchmark. The no-MS and common-profile channels are generated using the physical channel models developed in Secs.~\ref{sec:model_withoutMS} and~\ref{sec:model_withMS}, respectively, whereas the user-specific benchmark is constructed from the ideal curvature channels defined above.
The first and second rows of Fig.~\ref{fig:benchmark_performance} correspond, respectively, to the same-direction UE arrangement in Fig.~\ref{fig:multi_user_common_profile}(a) and the random-sector arrangement in Fig.~\ref{fig:multi_user_common_profile}(d). In both configurations, the common-profile MWB scheme improves the multiplexing capability relative to the no-MS case, although the gain remains limited by the use of a single shared MWB profile.
For the user-specific benchmark, the effective rank is equal to $K$, while the spectral efficiency sum increases approximately linearly with $K$ in both rows. These observations confirm that the benchmark provides mutually orthogonal channel vectors for all K users, regardless of the UE distribution. 
%

\pati{Physical Qualification}
Realizing the user-specific benchmark would require different curvature transformations to be applied simultaneously to the incident fields of different UEs. Such functionality cannot be achieved by a unique metasurface placed between the users and the BS. It would therefore require additional user-selective degrees of freedom, potentially enabled by a sophisticated constellation of active, nonlinear or spatiotemporally modulated metasurfaces.
Accordingly, the user-specific result here is interpreted as an idealized, upper-bound benchmark, illustrating the potential of future metasurface architectures with user-selective wavefront control. The following implementation discussion instead focuses on the common-profile MWB scheme, which is compatible with a passive transmission profile.

\section{Metasurface Implementation}
\label{sec:implementation}
\begin{figure}[!h]
    \centering
    \begin{subfigure}[t]{0.49\linewidth}
        \includegraphics[width=\linewidth]{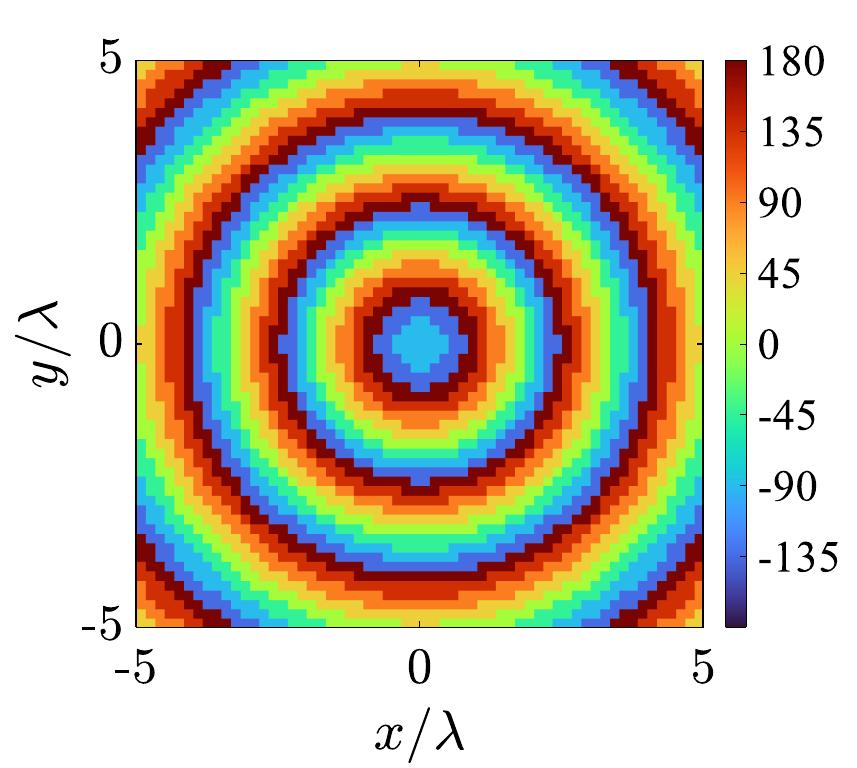}
        \captionsetup{skip=1pt}
        \caption{}
    \end{subfigure}
    \hfill
    \begin{subfigure}[t]{0.49\linewidth}
        \includegraphics[width=\linewidth]{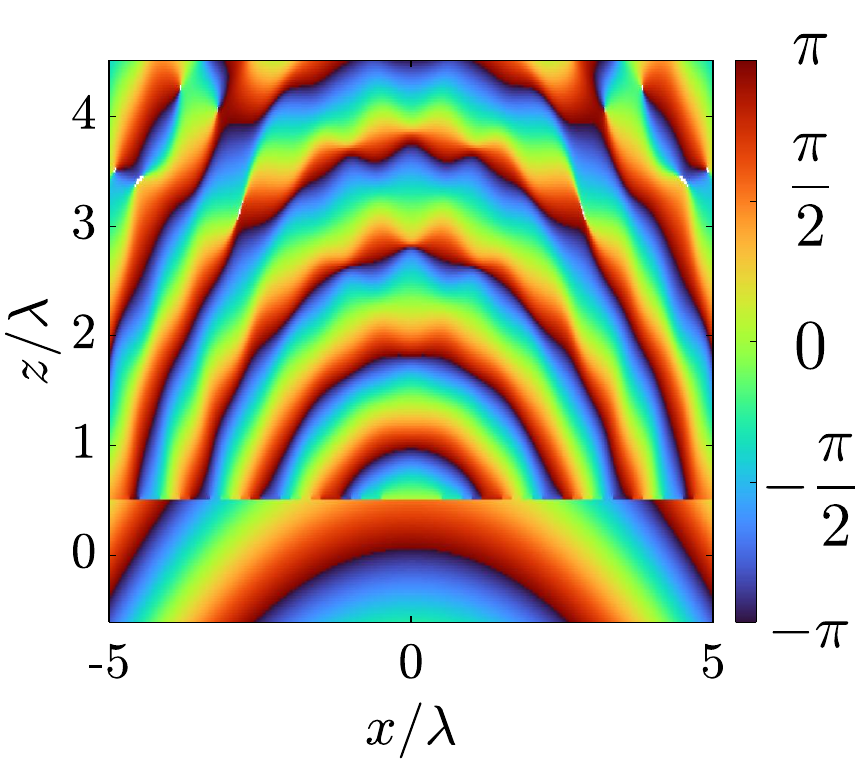}
        \captionsetup{skip=1pt}
        \caption{}
    \end{subfigure}
\caption{Eight-state quantized realization of the WFB metasurface. (a)~Transmission phase distribution obtained from discretizing the continuous profile in Fig.~\ref{fig:local_TR}(b) according to Eq.~\eqref{eq:phase_quantization_rule}. (b)~Resulting wavefront curvature bending transformation.}
\label{fig:dis_phase_bending}
\end{figure}

\pati{wavefront Bending under a Quantized Phase Distribution}
In the preceding sections, we have assumed a metasurface with a continuously varying local transmission phase associated with continuous mathematical susceptibility functions. In practice, however, a metasurface consists of finite-sized unit cells and can provide only a discrete set of transmission states. We therefore quantize here the continuous phase $\angle T(x,y)$ into eight uniformly spaced states over the full $2\pi$ range, corresponding to a phase resolution of $\Delta\phi=\pi/4$. The quantized transmission phase assigned to the unit cell at $(x_m,y_m)$ is therefore
\begin{equation}
\angle T_{\mathrm{quan}}(x_m,y_m)
=\Delta\phi\times\operatorname*{round}\left(\frac{\angle T(x_m,y_m)}{\Delta\phi}\right),
\label{eq:phase_quantization_rule}
\end{equation}
with the resulting phase wrapped into the interval $(-\pi,\pi]$. The corresponding local transmission coefficient is
$T_{\mathrm{quan}}(x_m,y_m)=\exp\left[j\angle T_{\mathrm{quan}}(x_m,y_m)\right]$.
Figure~\ref{fig:dis_phase_bending} shows the realization of the MWB transformation under $3$-bit phase quantization.
Figure~\ref{fig:dis_phase_bending}(a) shows the eight-state phase map obtained from the continuous GSTC-designed transmission phase in Fig.~\ref{fig:local_TR}(b), using Eq.~\eqref{eq:phase_quantization_rule}. Figure~\ref{fig:dis_phase_bending}(b) verifies the resulting transmitted field. The output wavefront remains more strongly curved than the incident field, indicating that the essential MWB operation survives finite phase quantization.
%

\begin{figure}[!h]
    \centering
    \begin{subfigure}[t]{0.49\linewidth}
        \includegraphics[width=\linewidth]{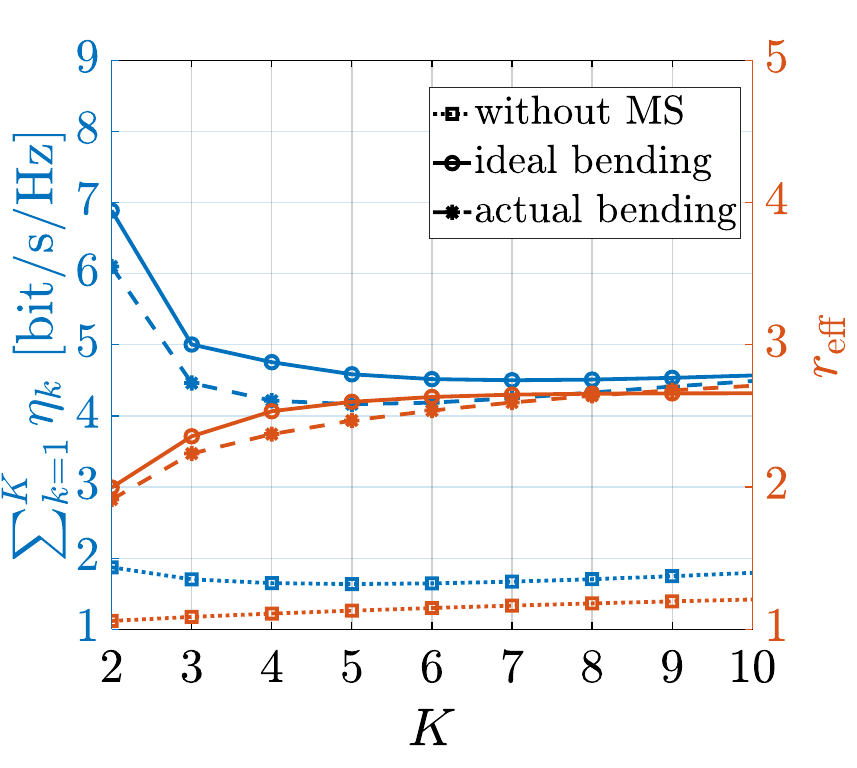}
        \captionsetup{skip=1pt}
        \caption{}
    \end{subfigure}
    \hfill
    \begin{subfigure}[t]{0.49\linewidth}
        \includegraphics[width=\linewidth]{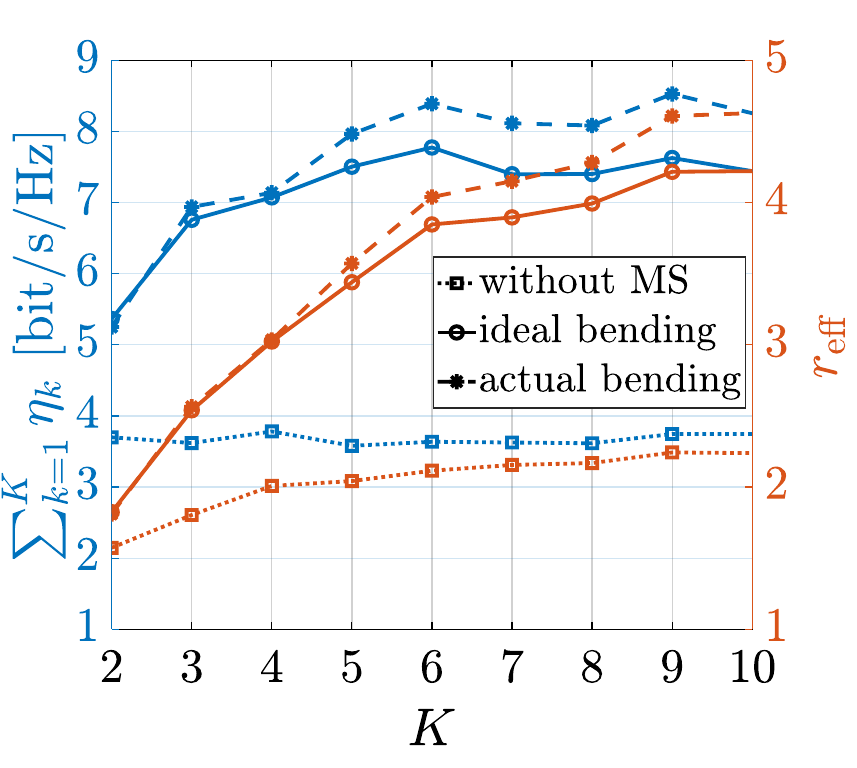}
        \captionsetup{skip=1pt}
        \caption{}
    \end{subfigure}
\caption{Effect of phase quantization on the common-profile MWB performance for $\beta=8$, with ideal-bending curves obtained for the continuous transmission phase in Fig.~\ref{fig:local_TR}(b) and actual-bending curves obtained after applying the discrete phase profile in Fig.~\ref{fig:dis_phase_bending}(a). The left and right vertical axes report the spectral efficiency sum computed from Eq.~\eqref{eq:SE} and the effective channel rank computed from Eq.~\eqref{eq:effective_rank}, respectively. (a)~Same-direction UE configuration in Fig.~\ref{fig:multi_user_common_profile}(a). (b)~Random-sector UE configuration in Fig.~\ref{fig:multi_user_common_profile}(d).}
\label{fig:ideal_vs_actual}
\end{figure}

\pati{MU-MIMO Performance under Actual MWB}
To evaluate the system-level impact of phase quantization, we replace the ideal common-profile transmission coefficient $T_{(i,\beta)}^{\mathrm{MS}}(x_m,y_m)$ in Eq.~\eqref{eq:complete_channel_dis} with the coefficient $T_{\mathrm{quan}}(x_m,y_m)$ constructed from Eq.~\eqref{eq:phase_quantization_rule}. We then assemble the channel vector of each UE using Eq.~\eqref{eq:complete_channel_dis}, and evaluate the spectral efficiency sum and effective channel rank using Eqs.~\eqref{eq:SE} and~\eqref{eq:effective_rank}, respectively. Figure~\ref{fig:ideal_vs_actual} compares the no-MS reference and the common-profile MWB scheme with $\beta=8$ using the ideal continuous phase profile in Fig.~\ref{fig:local_TR}(b) and the eight-state quantized profile in Fig.~\ref{fig:dis_phase_bending}(a). 
For both the same-direction UE configuration in Fig.~\ref{fig:multi_user_common_profile}(a) and the random-sector configuration in Fig.~\ref{fig:multi_user_common_profile}(d), phase quantization introduces only minor performance deviations while preserving most of the MWB gain relative to the no-MS case. These results demonstrate that $3$-bit phase control preserves the principal MU-MIMO multiplexing enhancement achieved by the ideal continuous-phase MWB system.
%

\begin{figure}[!h]
    \centering
    \begin{subfigure}[t]{0.49\linewidth}
        \includegraphics[width=\linewidth]{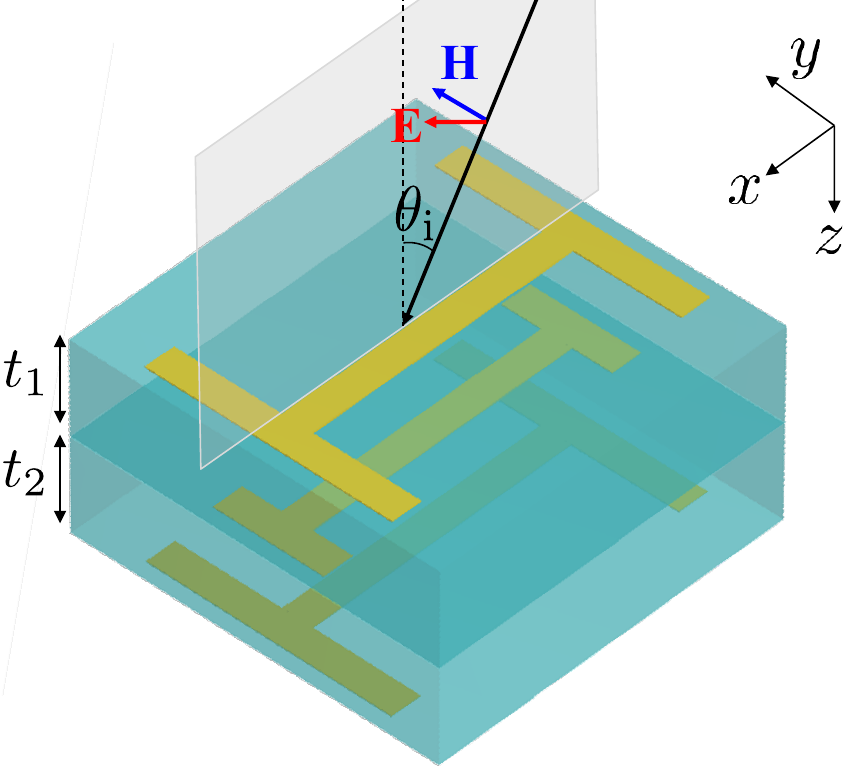}
        \captionsetup{skip=1pt}
        \caption{}
    \end{subfigure}
    \hfill
    \begin{subfigure}[t]{0.49\linewidth}
        \includegraphics[width=\linewidth]{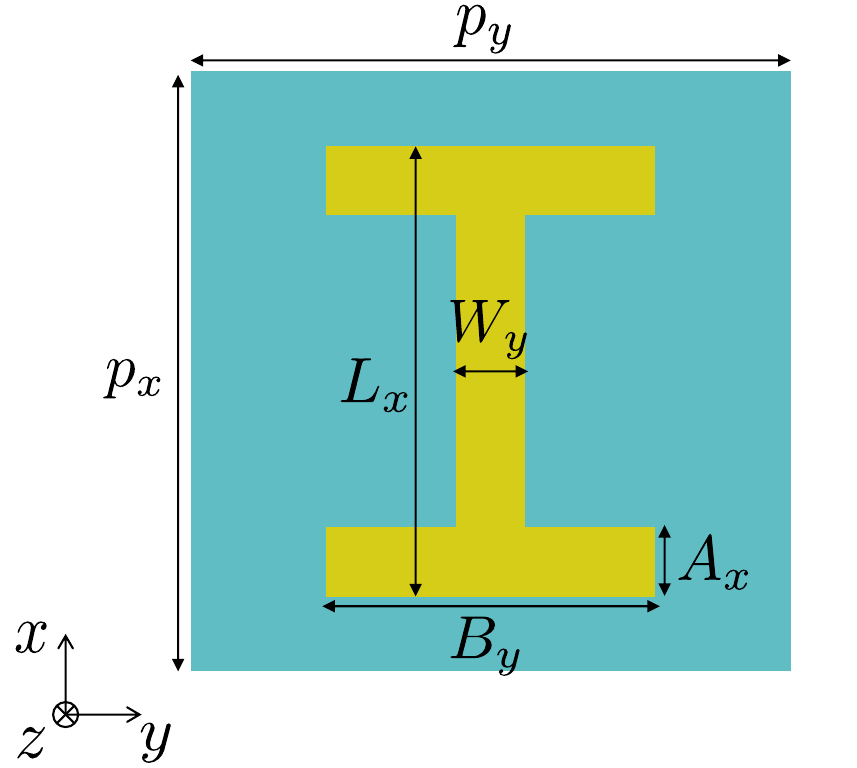}
        \captionsetup{skip=1pt}
        \caption{}
    \end{subfigure}
\caption{Metasurface scattering particle with generic variable dimensions. (a)~Unit cell perspective view with dielectric substrates made transparent for visualization. (b)~Unit cell front view with dogbone-shaped metallic particle.}
\label{fig:dog_bone_unitcell}
\end{figure}

\pati{Unit-Cell Geometry}
To realize a physical metasurface, we map the quantized phase states in Fig.~\ref{fig:dis_phase_bending}(a) onto subwavelength scattering particles and adopt three-layer dogbone Huygens metasurface architecture. The symmetric three-layer configuration provides the minimum degrees of freedom required to independently control the effective electric and magnetic surface responses. When these responses are properly balanced, the backward-scattered fields cancel, ideally yielding zero reflection and full transmission while providing complete $2\pi$ transmission-phase coverage~\cite{pfeiffer2013metamaterial,achouri2021electromagnetic}. The metallic layers are separated by two Rogers RO3003 dielectric spacers with relative permittivity $\epsilon_r=3$, loss tangent $\tan\delta=0.001$, and spacer thickness $1.52~\mathrm{mm}$, resulting in a total dielectric thickness of $3.04~\mathrm{mm}$. Figure~\ref{fig:dog_bone_unitcell} shows the adopted unit-cell topology. The metallic dogbone is oriented to accommodate the $x$-polarized incident field assumed in the channel model.  

\pati{Dimensions}
Based on the three-layer dog-bone architecture shown in Fig.~\ref{fig:dog_bone_unitcell}, eight unit cells are designed to realize the quantized transmission-phase states required by the MWB profile. Table~\ref{tab:unit_cell_dimensions} lists the optimized design parameters and the corresponding target transmission phase assigned to each unit cell.
\begin{table}[!h] \caption{Design parameters and target transmission phases for the eight unit cells. OL denotes the identical outer layers (top and bottom), and ML denotes the middle layer.}
\centering
\label{tab:unit_cell_dimensions}
\renewcommand{\arraystretch}{1.2}
\setlength{\tabcolsep}{8pt}
\begin{tabular}{c c c c c}
\toprule
\multicolumn{5}{c}{\textit{Common design parameters}} \\
\midrule
$f_0$ (GHz)
& \multicolumn{4}{c}{10} \\
$L_x^{\mathrm{OL}}=L_x^{\mathrm{ML}}$ (mm)
& \multicolumn{4}{c}{5.50} \\
$W_y^{\mathrm{OL}}=W_y^{\mathrm{ML}}$ (mm)
& \multicolumn{4}{c}{0.50} \\
$A_x^{\mathrm{OL}}=A_x^{\mathrm{ML}}$ (mm)
& \multicolumn{4}{c}{0.50} \\
$t_1=t_2$ (mm)
& \multicolumn{4}{c}{1.52} \\
$p_x=p_y$ (mm)
& \multicolumn{4}{c}{$\lambda_0/5=6.00$} \\
\addlinespace[5pt]

\midrule
& \text{Cell 1} & \text{Cell 2} & \text{Cell 3} & \text{Cell 4} \\
\midrule
Target phase
& $-135^\circ$ & $-90^\circ$ & $-45^\circ$ & $0^\circ$ \\
$B_y^{\mathrm{OL}}$ (mm)
& 1.77 & 1.26 & 0.70 & 5.80 \\
$B_y^{\mathrm{ML}}$ (mm)
& 4.50 & 5.80 & 4.50 & 1.24 \\
\addlinespace[5pt]

\midrule
& \text{Cell 5} & \text{Cell 6} & \text{Cell 7} & \text{Cell 8} \\
\midrule
Target phase
& $45^\circ$ & $90^\circ$ & $135^\circ$ & $180^\circ$ \\
$B_y^{\mathrm{OL}}$ (mm)
& 2.45 & 2.20 & 2.00 & 1.80 \\
$B_y^{\mathrm{ML}}$ (mm)
& 1.95 & 1.92 & 1.85 & 1.28 \\
\bottomrule
\end{tabular}
\end{table}
%

\begin{figure}[!h]
    \centering
    \begin{subfigure}[t]{0.99\linewidth}
        \includegraphics[width=\linewidth]{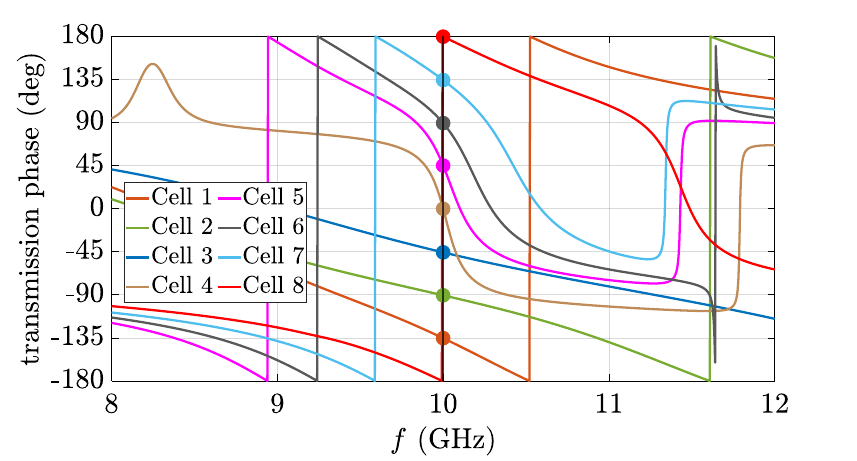}
        \captionsetup{skip=1pt}
        \caption{}
    \end{subfigure}
    \vfill
    \begin{subfigure}[t]{0.99\linewidth}
        \includegraphics[width=\linewidth]{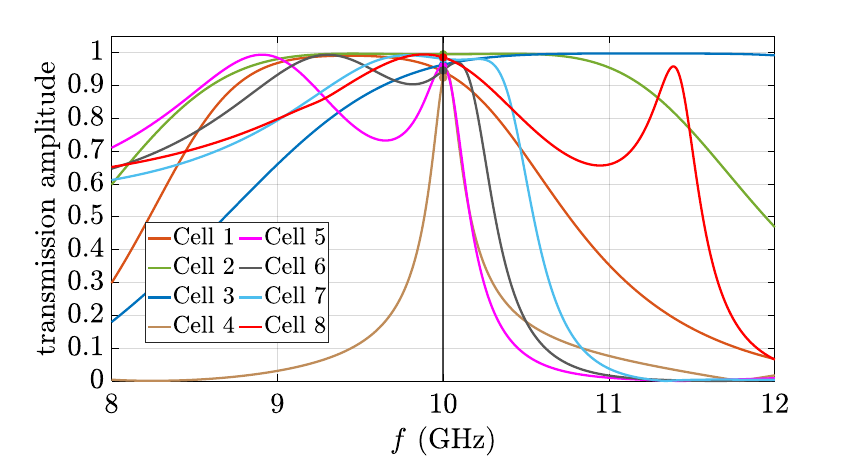}
        \captionsetup{skip=1pt}
        \caption{}
    \end{subfigure}
\caption{Simulated transmission responses of the eight three-layer dogbone unit cells under normal incidence, $\theta_i=0^\circ$. The unit-cell architecture is shown in Fig.~\ref{fig:dog_bone_unitcell}, and the corresponding geometrical parameters are listed in Table~\ref{tab:unit_cell_dimensions}. (a)~Transmission phase and (b)~transmission amplitude versus frequency.}
\label{fig:dog_bone_T_norm}
\end{figure}

\pati{Simulated Results}
We simulated the eight unit cells in CST Microwave Studio using the topology shown in Fig.~\ref{fig:dog_bone_unitcell}, the geometrical dimensions listed in Table~\ref{tab:unit_cell_dimensions} and periodic boundary conditions. 
Figure~\ref{fig:dog_bone_T_norm} presents the corresponding full-wave transmission responses under normal incidence, with $\theta_i=0^\circ$. Figure~\ref{fig:dog_bone_T_norm}(a) shows that the eight unit cells collectively cover the full $360^\circ$ transmission-phase range around the operating frequency of $10~\mathrm{GHz}$, with an approximately $45^\circ$ phase interval between adjacent states. Figure~\ref{fig:dog_bone_T_norm}(b) shows that most unit cells maintain high transmission magnitudes ($|T|>0.9$) near the operating frequency. These results confirm that the eight-state quantized phase profile in Fig.~\ref{fig:dis_phase_bending}(a) can be implemented using the proposed three-layer unit-cell library, thereby linking the GSTC-derived MWB phase profile to a realizable passive metasurface.
%


\section{Conclusion}
\label{sec:conclusion}
This paper introduced metasurface wavefront bending (MWB) to enhance radiative near-field MU-MIMO multiplexing by increasing spherical-wave curvature and reducing inter-user channel correlation. The proposed MWB design and system-level evaluations demonstrated substantial improvements in spectral efficiency sum and effective channel rank.  Future work will further optimize the metasurface unit-cell design to improve its robustness under oblique incidence and will focus on the fabrication and experimental validation of the proposed metasurface-assisted MU-MIMO system.

\begin{appendices}
\section{Derivation of the Correlation-Dependent SINR}
\label{app:sinr_correlation_derivation}

We consider two UEs, indexed by $i$ and $k$, with equal transmit powers $p_i=p_k=p$. To isolate spatial separability from channel path loss, we normalize the channel vectors as
\begin{equation}
\widetilde{\mathbf h}_i=\frac{\mathbf h_i}{\|\mathbf h_i\|},\qquad \widetilde{\mathbf h}_k=\frac{\mathbf h_k}{\|\mathbf h_k\|},
\label{eq:app_normalized_channels}
\end{equation}
such that $\|\widetilde{\mathbf h}_i\|=\|\widetilde{\mathbf h}_k\|=1$. For the detection of UE~$k$, we replace the channel vectors in Eq.~\eqref{eq:SINR_general} with the normalized ones in Eq.~\eqref{eq:app_normalized_channels}. The resulting normalized-channel SINR becomes
\begin{equation}
\overline{\mathrm{SINR}}_{k}=\frac{p\left|\mathbf v_k^{\mathrm H}\widetilde{\mathbf h}_k\right|^2}{p\left|\mathbf v_k^{\mathrm H}\widetilde{\mathbf h}_i\right|^2+\sigma^2\|\mathbf v_k\|^2}.
\label{eq:app_two_user_sinr}
\end{equation}
For this two-user system, the MMSE combining vector in Eq.~\eqref{eq:MMSE_combiner} becomes
\begin{equation}
\mathbf v_k^{\mathrm{MMSE}}=\left(p\widetilde{\mathbf h}_i\widetilde{\mathbf h}_i^{\mathrm H}+p\widetilde{\mathbf h}_k\widetilde{\mathbf h}_k^{\mathrm H}+\sigma^2\mathbf I_N\right)^{-1}\widetilde{\mathbf h}_k.
\label{eq:app_two_user_mmse}
\end{equation}
We define the interference-plus-noise terms in Eq.~\eqref{eq:app_two_user_mmse} as
\begin{equation}
\mathbf R_i\triangleq p\widetilde{\mathbf h}_i\widetilde{\mathbf h}_i^{\mathrm H}+\sigma^2\mathbf I_N.
\label{eq:app_interference_covariance}
\end{equation}
Substitution of Eq.~\eqref{eq:app_interference_covariance} into Eq.~\eqref{eq:app_two_user_mmse} gives
\begin{equation}
\mathbf v_k^{\mathrm{MMSE}}=\left(\mathbf R_i+p\widetilde{\mathbf h}_k\widetilde{\mathbf h}_k^{\mathrm H}\right)^{-1}\widetilde{\mathbf h}_k.
\label{eq:app_full_mmse}
\end{equation}
Applying the matrix inversion lemma to Eq.~\eqref{eq:app_full_mmse} yields
\begin{equation}
\mathbf v_k^{\mathrm{MMSE}}=\frac{\mathbf R_i^{-1}\widetilde{\mathbf h}_k}{1+p\widetilde{\mathbf h}_k^{\mathrm H}\mathbf R_i^{-1}\widetilde{\mathbf h}_k}.
\label{eq:app_mmse_scaled}
\end{equation}
The denominator in Eq.~\eqref{eq:app_mmse_scaled} is a nonzero scalar. Moreover, any nonzero scaling of the combining vector leaves the output SINR unchanged~\cite{tse2005fundamentals}:
\begin{equation}
\mathrm{SINR}_k(c\mathbf v_k)=\mathrm{SINR}_k(\mathbf v_k),\qquad c\neq 0.
\label{eq:app_scaling_invariance}
\end{equation}
Indeed, replacing $\mathbf v_k$ with $c\mathbf v_k$ in Eq.~\eqref{eq:app_two_user_sinr} multiplies the desired-signal, interference, and noise powers by the same factor $|c|^2$, which cancels between the numerator and denominator. Equation~\eqref{eq:app_scaling_invariance} therefore allows us to omit the scalar denominator in Eq.~\eqref{eq:app_mmse_scaled} and use the equivalent combining vector as
\begin{equation}
\mathbf v_k=\mathbf R_i^{-1}\widetilde{\mathbf h}_k.
\label{eq:app_equivalent_combiner}
\end{equation}

We first substitute Eq.~\eqref{eq:app_equivalent_combiner} into the numerator of Eq.~\eqref{eq:app_two_user_sinr} and obtain
\begin{equation}
p\left|\mathbf v_k^{\mathrm H}\widetilde{\mathbf h}_k\right|^2=p\left|\widetilde{\mathbf h}_k^{\mathrm H}\mathbf R_i^{-1}\widetilde{\mathbf h}_k\right|^2.
\label{eq:app_numerator_reduced}
\end{equation}
Using the definition of $\mathbf R_i$ in Eq.~\eqref{eq:app_interference_covariance}, we rewrite the denominator of Eq.~\eqref{eq:app_two_user_sinr} as
\begin{equation}
p\left|\mathbf v_k^{\mathrm H}\widetilde{\mathbf h}_i\right|^2+\sigma^2\|\mathbf v_k\|^2=\mathbf v_k^{\mathrm H}\left(p\widetilde{\mathbf h}_i\widetilde{\mathbf h}_i^{\mathrm H}+\sigma^2\mathbf I_N\right)\mathbf v_k=\mathbf v_k^{\mathrm H}\mathbf R_i\mathbf v_k.
\label{eq:app_denominator_quadratic}
\end{equation}
Substitution of Eq.~\eqref{eq:app_equivalent_combiner} into Eq.~\eqref{eq:app_denominator_quadratic} then gives
\begin{equation}
\mathbf v_k^{\mathrm H}\mathbf R_i\mathbf v_k=\widetilde{\mathbf h}_k^{\mathrm H}\mathbf R_i^{-1}\mathbf R_i\mathbf R_i^{-1}\widetilde{\mathbf h}_k=\widetilde{\mathbf h}_k^{\mathrm H}\mathbf R_i^{-1}\widetilde{\mathbf h}_k.
\label{eq:app_denominator_reduced}
\end{equation}
Because $\mathbf R_i$ is Hermitian positive definite, the quadratic form
$\widetilde{\mathbf h}_k^{\mathrm H}\mathbf R_i^{-1}\widetilde{\mathbf h}_k$
is real and strictly positive. We therefore substitute Eqs.~\eqref{eq:app_numerator_reduced} and~\eqref{eq:app_denominator_reduced} into Eq.~\eqref{eq:app_two_user_sinr} and cancel the common quadratic factor. This substitution yields
\begin{equation}
\overline{\mathrm{SINR}}_{k}=p\widetilde{\mathbf h}_k^{\mathrm H}\mathbf R_i^{-1}\widetilde{\mathbf h}_k.
\label{eq:app_mmse_sinr_compact}
\end{equation}
We next evaluate $\mathbf R_i^{-1}$. Applying the Sherman--Morrison identity~\cite{tse2005fundamentals} to Eq.~\eqref{eq:app_interference_covariance} gives
\begin{equation}
\mathbf R_i^{-1}=\left(\sigma^2\mathbf I_N+p\widetilde{\mathbf h}_i\widetilde{\mathbf h}_i^{\mathrm H}\right)^{-1}=\frac{1}{\sigma^2}\left[\mathbf I_N-\frac{p}{\sigma^2+p\|\widetilde{\mathbf h}_i\|^2}\widetilde{\mathbf h}_i\widetilde{\mathbf h}_i^{\mathrm H}\right].
\label{eq:app_sherman_morrison_general}
\end{equation}
Using $\|\widetilde{\mathbf h}_i\|^2=1$, we simplify Eq.~\eqref{eq:app_sherman_morrison_general} to
\begin{equation}
\mathbf R_i^{-1}=\frac{1}{\sigma^2}\left[\mathbf I_N-\frac{p}{\sigma^2+p}\widetilde{\mathbf h}_i\widetilde{\mathbf h}_i^{\mathrm H}\right].
\label{eq:app_covariance_inverse}
\end{equation}
Substituting Eq.~\eqref{eq:app_covariance_inverse} into Eq.~\eqref{eq:app_mmse_sinr_compact} yields
\begin{align}
\overline{\mathrm{SINR}}_{k}&=\frac{p}{\sigma^2}\widetilde{\mathbf h}_k^{\mathrm H}\left[\mathbf I_N-\frac{p}{\sigma^2+p}\widetilde{\mathbf h}_i\widetilde{\mathbf h}_i^{\mathrm H}\right]\widetilde{\mathbf h}_k \nonumber \\
&=\frac{p}{\sigma^2}\left[\widetilde{\mathbf h}_k^{\mathrm H}\widetilde{\mathbf h}_k-\frac{p}{\sigma^2+p}\widetilde{\mathbf h}_k^{\mathrm H}\widetilde{\mathbf h}_i\widetilde{\mathbf h}_i^{\mathrm H}\widetilde{\mathbf h}_k\right],
\label{eq:app_sinr_substitution}
\end{align}
where the first term satisfies $\widetilde{\mathbf h}_k^{\mathrm H}\widetilde{\mathbf h}_k=1$, 
and the term can be written as $\widetilde{\mathbf h}_k^{\mathrm H}\widetilde{\mathbf h}_i\widetilde{\mathbf h}_i^{\mathrm H}\widetilde{\mathbf h}_k=\left|\widetilde{\mathbf h}_i^{\mathrm H}\widetilde{\mathbf h}_k\right|^2=\bar{\rho}_{ik}^{\,2}$.
Consequently, Eq.~\eqref{eq:app_sinr_substitution} becomes
\begin{equation}
\overline{\mathrm{SINR}}_{k}=\frac{p}{\sigma^2}\left(1-\frac{p}{\sigma^2+p}\bar{\rho}_{ik}^{\,2}\right).
\label{eq:app_sinr_power_form}
\end{equation}
Finally, defining the interference-free reference SNR as
\begin{equation}
\gamma\triangleq\frac{p}{\sigma^2},
\label{eq:app_gamma_definition}
\end{equation}
and substituting Eq.\eqref{eq:app_gamma_definition} into Eq.\eqref{eq:app_sinr_power_form} gives the desired correlation-dependent SINR expression
\begin{equation}
\overline{\mathrm{SINR}}_{k}=\gamma\left(1-\frac{\gamma}{1+\gamma}\bar{\rho}_{ik}^{\,2}\right),
\label{eq:app_final_sinr}
\end{equation}
which explicitly separates the effect of the interference-free reference SNR from that of the normalized channel correlation between the two users.

\section{Selected Bending Ratios for the User-Specific MWB Benchmark}
\label{app:selected_user_specific_beta}

This appendix reports the user-specific bending-ratio sets obtained from the sequential minimax-correlation selection in Eq.~\eqref{eq:user_specific_beta_selection}.
The selected bending ratios are nonuniformly distributed over the available range. This behavior results from the nonlinear dependence of the channel correlation on the bending ratio and from the combined effects of the UE geometry, finite metasurface and BS apertures, and UE--metasurface--BS propagation. As $K$ increases, the previously selected values are retained and one additional bending ratio is introduced, yielding the nested solution sets listed in Table~\ref{tab:selected_user_specific_beta}.
\begin{table}[h!]
\centering
\caption{Selected bending ratios for the user-specific MWB benchmark in the random-sector UE configuration.}
\label{tab:selected_user_specific_beta}
\renewcommand{\arraystretch}{1.15}
\begin{tabular}{c l}
\hline
$K$ & Selected bending ratios $\{\beta_k\}_{k=1}^{K}$ \\
\hline
$2$  & $\{1.00,\;3.00\}$ \\
$3$  & $\{1.00,\;3.00,\;8.00\}$ \\
$4$  & $\{1.00,\;3.00,\;4.27,\;8.00\}$ \\
$5$  & $\{1.00,\;3.00,\;4.27,\;8.00,\;9.36\}$ \\
$6$  & $\{1.00,\;1.73,\;3.00,\;4.27,\;8.00,\;9.36\}$ \\
$7$  & $\{1.00,\;1.73,\;3.00,\;4.27,\;6.09,\;8.00,\;9.36\}$ \\
$8$  & $\{1.00,\;1.73,\;3.00,\;4.27,\;6.09,\;6.82,\;8.00,\;9.36\}$ \\
$9$  & $\{1.00,\;1.73,\;3.00,\;4.27,\;6.09,\;6.82,\;7.27,\;8.00,\;9.36\}$ \\
$10$ & $\{1.00,\;1.73,\;3.00,\;4.27,\;4.64,\;6.09,\;6.82,\;7.27,\;8.00,\;9.36\}$ \\
\hline
\end{tabular}
\end{table}

\end{appendices}

\bibliographystyle{IEEEtran}
\bibliography{MS_Curvature_MIMO}

\end{document}